\documentclass[twocolumn,english,prc,reprint]{revtex4}
\usepackage[T1]{fontenc}
\usepackage[latin9]{inputenc}
\usepackage{textcomp}
\usepackage{amsmath}
\usepackage{amssymb}
\usepackage{graphicx}

\usepackage{bbold}

\makeatletter

\newcommand{\lyxmathsym}[1]{\ifmmode\begingroup\def\b@ld{bold}
  \text{\ifx\math@version\b@ld\bfseries\fi#1}\endgroup\else#1\fi}

\providecommand{\tabularnewline}{\\}

\@ifundefined{textcolor}{}
{%
 \definecolor{BLACK}{gray}{0}
 \definecolor{WHITE}{gray}{1}
 \definecolor{RED}{rgb}{1,0,0}
 \definecolor{GREEN}{rgb}{0,1,0}
 \definecolor{BLUE}{rgb}{0,0,1}
 \definecolor{CYAN}{cmyk}{1,0,0,0}
 \definecolor{MAGENTA}{cmyk}{0,1,0,0}
 \definecolor{YELLOW}{cmyk}{0,0,1,0}
}

\usepackage{physics}

\usepackage{babel}
\newcommand{\pr}[1]{{\sc{\lowercase{#1}}}}

\makeatother

\usepackage{babel}
\begin{document}
\title{Nuclear spectra from low-energy interactions}
\author{J. Ljungberg, B. G. Carlsson, J. Rotureau, A. Idini and I. Ragnarsson}
\affiliation{Mathematical Physics, LTH, Lund University, S-22100 Lund, Sweden}
\date{\today}
\begin{abstract}
A method to describe spectra starting from nuclear density functionals
is explored. The idea is based on postulating an effective Hamiltonian
that reproduces the stiffness associated with collective modes. The
method defines a simple form of such an effective Hamiltonian and
a mapping to go from a density functional to the corresponding Hamiltonian.
In order to test the method, the Hamiltonian is constrained using
a Skyrme functional and solved with the generator-coordinate method
to describe low-lying levels and electromagnetic transitions in $^{48,49,50,52}$Cr
and $^{24}$Mg. 
\end{abstract}
\maketitle

\section{Introduction}

A starting point for the description of nuclei is the assumption that
low-energy properties can be described using a combination of two-
and three-body interactions. This sought after Hamiltonian should
be applicable to all nuclei and give reliable predictions for properties
that are not yet measured. A possible route for finding such an interaction
comes from Skyrme's expansion in the relative momenta of interacting
nucleons \citep{SKYRME1958}. This expansion can be carried out to
higher orders \citep{carlsson2008-n3,raimondi2011} and recently the
first applications of such higher order interactions has emerged \citep{ryssens2021}. 

For essentially all applications, such as descriptions of fission
and for systematic descriptions of nuclei and reactions, the method
has to be numerically efficient in order to be useful. This leads
to approximations where for example finite-range three-body interactions
can not be treated explicitly and the three-body part is conveniently
described as a density-dependent two-body interaction. The resulting
approximations are known as nuclear energy density functionals \citep{Bender2003}
(EDF's). When used in connection with Hartree-Fock-Bogoliubov (HFB)
approximations they describe ground-state masses with an error of
around 0.7 MeV \citep{scamps2021}. When used in other approaches
such as the quasiparticle-random-phase approximation (QRPA) they describe
many observables such as low-lying excitations and strength functions
\citep{vesely12,QRPA-low-lying}. This indicates that the original
assumption of a common low-energy interaction applicable to all nuclei
is not that far fetched. 

Nuclei have a tendency towards spontaneous symmetry breaking, in
particular pertaining to their shapes. Therefore, in order to capture
physical effects the description with EDF's is based on the breaking
of symmetries. This leads to intuitive and rather accurate descriptions
in terms of deformed nuclei with broken quantum numbers. However,
in an exact treatment, the nuclear wave functions should be eigenstates
of operators corresponding to conserved quantities, such as the squared
total angular momentum operator. Methods that restore the symmetries
give corrections to binding energies and allow a direct comparison
of observables, such as energy levels, to experiment. Restoration
of symmetries can be done in several ways but a common theme is to
reduce the degrees of freedom of the system in order to keep the efficiency
and applicability of the methods. 

Several such approaches have been developed that do not introduce
any free parameters but rather determine the parameters from the response
of the EDF's to external fields. Examples of such approaches includes;
the particle(s)-rotor model \citep{MPR}, where the system is divided
into a collective rotor part and a particle part; the Bohr Hamiltonian,
where also vibrations of the shape of the rotor is included \citep{Bohr2010};
the interacting boson-fermion model \citep{Nomura2020}, where the
degrees of freedom are mapped into interacting effective particles;
as well as methods to construct effective simpler Hamiltonians from
underlying EDF's \citep{Bertsch2006}. 

One of the most promising directions for symmetry restoration is based
on the generating coordinate method (GCM) where the problem of choosing
degrees of freedom is converted into choosing an appropriate subspace
of non-orthogonal many-body basis states. The degrees of freedom are
instead selected by choosing external fields for sampling of the space.
Its microscopic nature and the possibilities for systematic convergence
are part of the appealing features of the GCM. In principle it can
be applied using a full low-energy interaction with finite range two-
and three body terms. However, to construct an efficient and applicable
approach it would be very convenient to be able to apply it together
with the already developed EDF's. 

In this respect, one issue is that models for the low-momentum part
of the interaction also contain a high-momentum part that may not
give physical results if it is not constrained. In Hartree-Fock type
of calculations, the high-momentum part of the interaction is never
probed. However, extensions that attempt to sum up correlation energies
may require momentum cutoffs in order to avoid ultraviolet divergencies
\citep{vonbarth2013}. A second issue is how to treat the density-dependent
part of the interaction. There are various approaches for obtaining
approximate matrix elements between many-body states that should represent
the physics contained in the density-dependent part of the interaction.
However, a difficulty in finding consistent approaches is that approximations
that violate the Pauli principle lead to poles that can cause nonphysical
contributions to the energy \citep{fritz1998,poles2009}.

In this article we use a Skyrme based EDF to constrain a simple effective
Hamiltonian that is based on the fundamental nuclear degrees of freedom
of quadrupole deformation and pairing. The Hamiltonian may be considered
to be composed of the first terms in a serie where degrees of freedom
are chosen through the selected multipole operators and the precision
of the expansion is determined by the number of terms included. This
Hamiltonian is used in GCM calculations, restoring the broken symmetries,
to obtain the ground state binding energies, nuclear spectra, and
transitions for even and odd nuclei. We take particular care to include
all exchange terms in order to avoid any spurious pole contributions
and make use of recent developments in the calculation of overlaps
of Bogoliubov states \citep{car21}. We recently applied the same
approach to describe excitations in superheavy nuclei \citep{samark2021}.
Here we provide a more detailed description of the formalism and present
results for several lighter nuclei, including electromagnetic transition
probabilities.

In sec. II we detail the structure of the Hamiltonian and the procedure
to link it to the EDF. In sec III we apply the method to several nuclei
and compare spectra and transitions with experiment. In sec. IV we
summarize our conclusions from the study. Further details on the many-body
formalism are given in the appendix.

\section{Model\label{sec:Model}}

\subsection{Effective Hamiltonian}

The starting point is the definition of an effective Hamiltonian.
This will eventually be solved in a basis of HFB states using the
GCM approach. In order to have an efficient and applicable method,
the Hamiltonian is chosen as: 
\begin{equation}
\hat{H}=\hat{H}_{0}+\hat{H}_{Q}+\hat{H}_{P}.\label{eq:H_m}
\end{equation}
$H$ includes three components to capture the most important physical
effects: a spherical single--particle (s.p.) potential $H_{0}$ that
averages the interaction among nucleus; a quadrupole-quadrupole interaction
$H_{Q}$ that takes into account the quadrupole deformation; and finally,
a pairing term $H_{P}$ to consider neutron--neutron and proton--proton
pairing correlations.

The s.p. potential is written as 
\begin{equation}
\hat{H}_{0}=\sum_{i}e_{i}a_{i}^{\dagger}a_{i}+E_{0},
\end{equation}
where $i\equiv(q_{i}n_{i}l_{i}j_{i}m_{i})$ denotes an orbital in
a spherical basis labeled with its particle species $q_{i}$ (=$p$
or $n$), principal quantum number $n_{i}$, angular momentum $l_{i}$,
total angular momentum $j_{i}$ and its projection $m_{i}$. The $e_{i}$
are the single-particle energies and $E_{0}$ is a constant. For convenience,
a separable form is chosen for both $\hat{H}_{Q}$ and $\hat{H}_{P}$.

The quadrupole--quadrupole separable interaction is given by
\begin{equation}
\hat{H}_{Q}=-\frac{1}{4}\chi\sum_{ijkl}\sum_{\mu}\left[\tilde{Q}_{ik}^{2\mu}\tilde{Q}_{lj}^{2\mu*}-\tilde{Q}_{il}^{2\mu}\tilde{Q}_{kj}^{2\mu*}\right]a_{i}^{\dagger}a_{j}^{\dagger}a_{l}a_{k},
\end{equation}
where $\chi$ is the interaction strength and $\tilde{Q}_{ij}^{2\mu}$
are the matrix elements of a modified quadrupole operator  with a
radial form factor. In this case, the form factor is based on a Woods-Saxon
potential from \citep{KUMAR1970} ( cf. Sec. \ref{sec:map}).

For the pairing part we adopt the seniority pairing interaction \citep{nilsson1995}
\begin{equation}
\hat{H}_{P}=-\frac{1}{4}\sum_{ijkl}G_{ik}P_{ij}P_{kl}a_{i}^{\dagger}a_{j}^{\dagger}a_{l}a_{k},\label{eq:Hp}
\end{equation}
where $G_{ik}=G_{p}\delta_{q_{i},p}\delta_{q_{k}p}+G_{n}\delta_{q_{i},n}\delta_{q_{k},n}$
is the pairing strength and 
\begin{equation}
P_{ij}=\left(-1\right)^{j_{i}-m_{i}}\delta_{\left(qnlj\right)_{i},\left(qnlj\right)_{j}}\delta_{m_{i},-m_{j}},
\end{equation}
indicates the coupling of time--reversal pairs only. The seniority
pairing is the simplest form of pairing interaction which nonetheless
enables a quantitative account of pairing phenomena and many-body
correlations \citep{Idini:12,Potel:17,Potel:21}. We fix the pairing
strength $G$ according to the uniform spectra method \citep{nilsson1995}
(see Sec. \ref{sec:map}). 

The resulting Hamiltonian contains the monopole, pairing and quadrupole
components. These are the well known dominant contributions responsible
e.g. for the behavior of isotopic chains and the shell evolution until
the drip lines \citet{tsunoda2020impact}. The Hamiltonian preserves
symmetries, such as exchange, rotational invariance and parity. Isospin
is violated by the quadrupole interaction that has a Coulomb part
in the form factor. The translational symmetry is also broken both
by the introduction of a single--particle potential, and by the decomposition
of the interaction into a finite number of separable terms. This is
however the case with any interaction represented on a grid of basis
functions.

\subsection{Determination of coupling constants \label{sec:map}}

The $H_{0}$ part of the Hamiltonian is taken as the spherical Hartree-Fock
(HF) potential from a Skyrme functional that will be the reference
for the effective Hamiltonian. The constant $E_{0}$ is taken to reproduce
the corresponding spherical HF binding energy. The quadrupole part
is also constructed to agree with the Skyrme results.

For neutrons, the quadrupole operator in $\hat{H}_{Q}$ is taken from
the modified quadrupole force in \citep{KUMAR1970}, 
\begin{equation}
\tilde{Q}_{ij}^{2\mu}=\left\langle i\left|\tilde{Q}_{p}^{2\mu}\right|j\right\rangle \delta_{q_{i},p}\delta_{q_{j},p}+\left\langle i\left|\tilde{Q}_{n}^{2\mu}\right|j\right\rangle \delta_{q_{i},n}\delta_{q_{j},n}\;,
\end{equation}
 where 

\begin{align}
\tilde{Q}_{n}^{2\mu}= & Y_{2\mu}\nonumber \\
 & \times\left(-R_{n}W_{n}\frac{\partial f_{n}\left(r\right)}{\partial r}+\frac{W_{n}v_{so}\lambda^{2}}{2}\frac{\partial^{2}f_{n}\left(r\right)}{\partial r^{2}}\vec{l}\cdot\vec{s}\right),
\end{align}
and 
\begin{equation}
f_{n}\left(r\right)=\frac{1}{1+e^{\left(r-R_{n}\right)/a}}.
\end{equation}
The proton part of the quadrupole operator

\begin{align}
\hat{Q}_{p}^{2\mu}= & Y_{2\mu}\left(-r\frac{\partial H_{c}}{\partial r}\right.\\
 & \left.-R_{p}W_{p}\frac{\partial f_{p}\left(r\right)}{\partial r}+\frac{W_{p}v_{so}\lambda^{2}}{2}\frac{\partial^{2}f_{p}\left(r\right)}{\partial r^{2}}\vec{l}\cdot\vec{s}\right),
\end{align}
has an additional dependence on the Coulomb potential 
\begin{align}
H_{c}= & \frac{Ze^{2}}{4\pi\epsilon_{0}}\left(\frac{1}{r}\theta\left(r-R_{p}\right)\right.\nonumber \\
 & \left.+\frac{1}{R_{p}}\left(\frac{3}{2}-\frac{1}{2}\left(\frac{r}{R_{p}}\right)^{2}\theta\left(R_{p}-r\right)\right)\right).
\end{align}

The quadrupole operators depend on the two radius parameters $R_{p}$
and $R_{n}$ for the proton and neutron densities. These are determined
from the expectation value of $r^{2}$ calculated from the spherical
Hartree--Fock solutions of the reference functional
\begin{table}
\begin{tabular}{cl}
\hline 
Quantity & Definition\tabularnewline
\hline 
$R_{q}$ & $=0.9\sqrt{\frac{5}{3}\left\langle r^{2}\right\rangle _{q}}$\tabularnewline
$a$ & $=0.9$ fm\tabularnewline
$W^{p/n}$ & $=V_{0}\left(1\pm\kappa\frac{N-Z}{N+Z}\right)$\tabularnewline
$V_{0}$ & $=-49.6$ MeV\tabularnewline
$\kappa$ & $=0.86$\tabularnewline
$v_{s.o.}$ & $=32$ $($MeV s$)^{-2}$\tabularnewline
$\lambda$ & $=\frac{\hbar}{Mc}\left(1+A^{-1}\right)$\tabularnewline
$M$ & $=939$ MeV/$c^{2}$\tabularnewline
\end{tabular}

\caption{Parameters defining the quadrupole interaction. For $W$ the upper
sign is associated with protons and lower sign with neutrons. \label{fig:Parameters-qq}}
\end{table}
. All the parameters of the interaction are in Tab.~\ref{fig:Parameters-qq}.
We keep the spin-orbit strength of \citep{KUMAR1970} but we use Universal
parametrization \citep{UniversalWS} for the other values. The diffuseness
constant is taken to be larger than in \citep{UniversalWS} since
in our initial tests we found that a larger value generally gives
more accurate reproduction of the EDF energy as a function of deformation.

\begin{figure}
\includegraphics[clip,width=0.6\columnwidth]{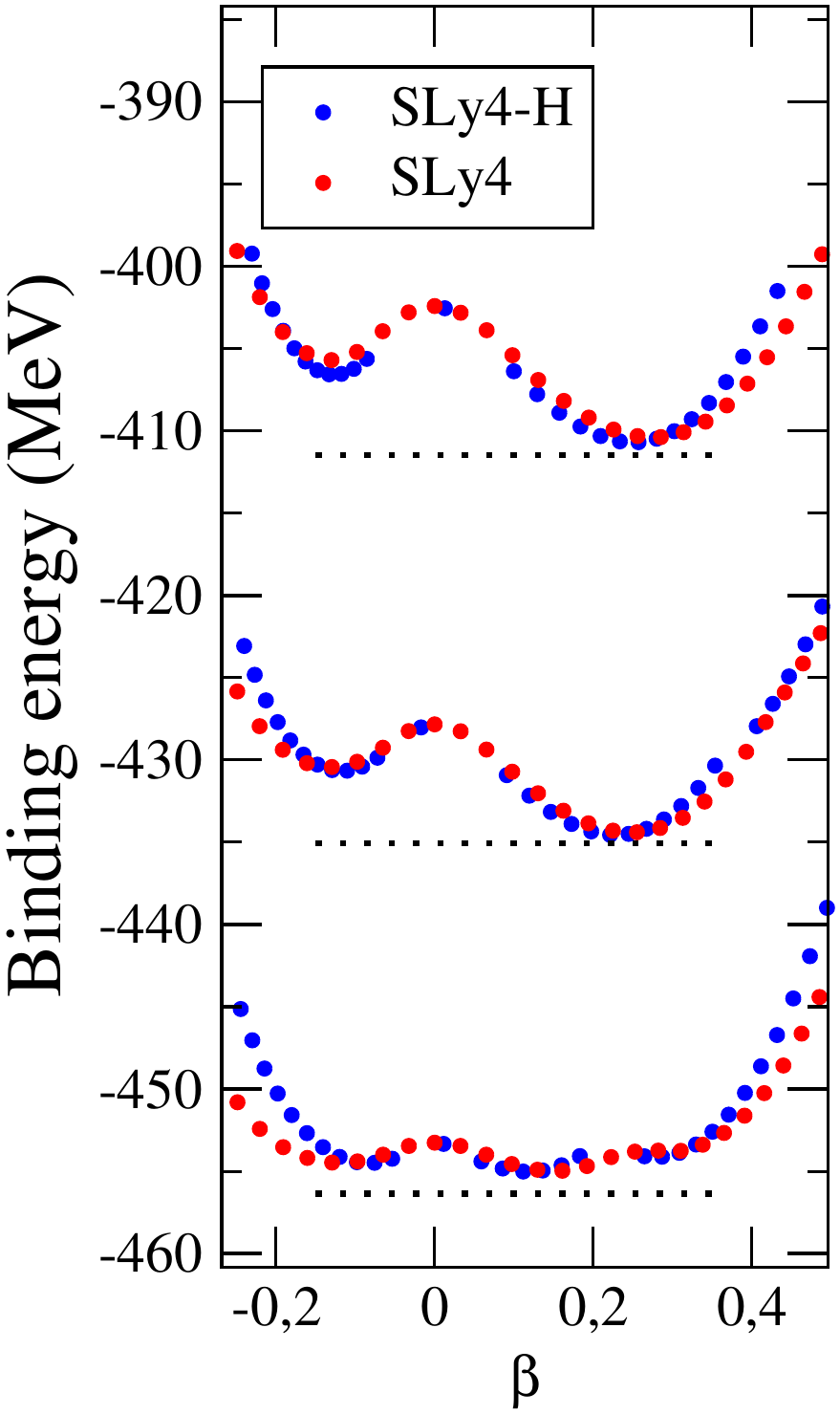}\caption{HF energy versus deformation $\beta$ for $^{48,50,52}$Cr using a
basis of 12 spherical oscillator shells ($N_{max}=11$). Experimental
binding energies are shown with dashed lines. The SLy4 EDF results
are obtained with the code \citep{HFBTHO} and the results abbreviated
SLy4-H are obtained with the effective Hamiltonian (\ref{eq:H_m})\label{fig:HF-energy-versus}}
\end{figure}

The strength of the quadrupole interaction $\chi$ is determined by
fitting the cost of deforming. Thus we fix the quadrupole-quadrupole
strength $\chi$ in the following manner: (i) for several values of
the deformation parameter $\beta_{2}$ we compute the energy $E^{HF}(\beta_{2})$
obtained with the functional within a Skyrme HF calculation with constraints
on the quadrupole moment, (ii) $\chi$ is then fitted such that the
HF energies obtained with the effective Hamiltonian $\hat{H}$ (\ref{eq:H_m}),
reproduces $E^{HF}(\beta_{2})$. As seen in Fig.~\ref{fig:HF-energy-versus}
the cost of deforming can be reproduced in a reasonable way. The approximation
of only having a single quadrupole term limits the range of deformations
that can be described. Thus the agreement is expected to deteriorate
for larger deformations where more complex shapes become important. 

For the pairing part of the Hamiltonian $H_{P}$ (Eq.~(\ref{eq:Hp}))
the interaction strength is determined applying the uniform model
of \citep{nilsson1995}. This treatment of pairing is similar to the
one in the Cranked-Nilsson-Strutinsky-Bogoliubov model (CNSB) \citep{carlsson2008}.
First we assume the empirical estimate of the average gap, $\tilde{\Delta}_{0}=0.7\times12/\sqrt{A}$
\citep{bohr1969nuclearvol1}. The reduction factor of 0.7 comes from
compensating for the effect of particle--number projection \citep{OLOFSSON2007,Pairing0.7}.
Then, the strength $G_{q}(\tilde{\Delta}_{0})$ is found by solving
the uniform model \citep{nilsson1995}. For each nucleus we obtain
different strengths $G_{q}$ by solving separately for protons and
neutrons,
\begin{equation}
\left(\tilde{\Delta}_{0}-G_{q}\right)=2Se^{-\frac{1}{G_{q}\rho}}\;,
\end{equation}
where the pairing window is set to $S=30$ MeV in the spherical basis.
Here $\rho$ denotes the level density averaged in the energy window
taken from the spherical HF solution. Note that the left hand side
is modified to take into account the contribution of the exchange
term in the pairing. The full treatment of the exchange term is needed
to avoid the singularities when applying the projection operator and
gives an extra contribution when breaking a pair (see appendix of
\citep{carlsson2008}).

In the way described in this section, all the coupling constants of
the effective Hamiltonian becomes determined from the underlying reference
functional. In this article we consistently apply the SLy4 parametrization
of the Skyrme interaction \citep{CHABANAT1998} as a reference functional
and denote the resulting Hamiltonian SLy4-H.

\subsection{Collective coordinates}

\label{sec:mb} In the previous sections, we defined the effective
Hamiltonian and its parameters. In the following, we define the many-body
basis within which the Hamiltonian is solved. Our basis states consists
of HFB vacua obtained with the effective Hamiltonian for several values
of the deformation located on a grid. This grid is constructed by
solving the HFB equations for the Hamiltonian in Eq.~(\ref{eq:H_m})
with constraints on,

\begin{align}
\beta_{x} & =\frac{4\pi}{5}\frac{\left\langle \hat{Q}_{20}\right\rangle }{\left\langle r^{2}\right\rangle },\\
\beta_{y} & =\frac{4\pi}{5}\sqrt{2}\frac{\left\langle \hat{Q}_{22}+\hat{Q}_{2-2}\right\rangle }{\left\langle r^{2}\right\rangle },
\end{align}
 with $\left\langle \bullet\right\rangle $ the expectation value
of the operator respect to the deformed HFB states and $\hat{Q}_{2\mu}=r^{2}Y_{2\mu}$.
From this, one obtains the familiar $\beta=\sqrt{\beta_{x}^{2}+\beta_{y}^{2}}$,
which defines the degree of quadrupole deformation, and $\gamma=\arctan\left(\frac{\beta_{y}}{\beta_{x}}\right)$,
which defines the trixiality. We also use the cranking method with
a constraint on,
\begin{equation}
j_{x}=\left\langle \hat{j}_{x}\right\rangle .
\end{equation}
In addition, we also include a variation of the pairing strengths
$G_{p}(\tilde{\Delta}_{0})$ and $G_{n}(\tilde{\Delta}_{0})$ by scaling
the pairing gaps $\tilde{\Delta}_{0}=g_{q}\Delta_{0}$. The many-body
basis states are obtained as the lowest energy solutions to the HFB
equations in a grid of $\beta,\,\gamma,\,j_{x},\,g_{p}$, and $g_{n}$
values. The grids are generated by sampling a region of the $\left(\beta,\gamma\right)$
plane. Each point of the plane can be associated with a certain value
of $(j_{x},\,g_{p},\,g_{n})$. We have allowed a few different values
of each of these variables and randomly assigned one of these values
for each $\left(\beta,\gamma\right)$ point. Only HFB states below
a certain cut-off energy are kept and accepted as basis states. 

This choice of generating coordinates attempts to account for the
most important collective degrees of freedom namely: collective vibrations
in the quadrupole degrees of freedom, rotations and pairing correlations.
In order to improve the accuracy for a larger class of states in the
spectrum one would need to enlarge the basis further by, for instance,
including states built through quasiparticle (qp.) excitations. An
equivalent way of introducing such non-collective particle-type excitations
that is more in the spirit of the GCM is to act on the basis states
with an excitation operator: 
\begin{align}
\left|\phi_{1}\right\rangle  & =\mathcal{N}e^{\hat{Z}}\left|\phi_{0}\right\rangle \nonumber \\
 & =\mathcal{N}\left(1+\hat{Z}+\frac{1}{2}\hat{Z}^{2}+...\right)\left|\phi_{0}\right\rangle ,\label{excop}
\end{align}
where $\mathcal{N}$ is a normalization constant and $\hat{Z}$ is
a two-quasiparticle creation operator:

\begin{equation}
\hat{Z}=\sum_{k<k'}z_{k,k'}\beta_{k}^{\dagger}\beta_{k'}^{\dagger}.\label{eq:qp-pairs}
\end{equation}
The $z_{k,k'}$ elements are chosen as

\begin{equation}
z_{k,k'}=e^{-\left(E_{k}+E_{k'}\right)/\left(k_{B}T\right)}\times p,\qquad\textrm{{for}\;}k'>k,
\end{equation}
with $z_{k,k'}=-z_{k',k}$, for $k'<k$ to ensure symmetry. For each
many-body state and for each matrix element, $p$ is randomly taken
as $\pm1$. The value of $k_{B}T$ is obtained from a parameter $b$
as, 
\begin{equation}
k_{B}T=-\left(E_{1}+E_{2}\right)/\ln\left(b\right).\label{kt}
\end{equation}
The smallest value of the sum of the lowest qp. energies; $E_{1}+E_{2}$
for either protons or neutrons are used for both particle species
in this relation.
\begin{figure}
\includegraphics[clip,width=0.95\columnwidth]{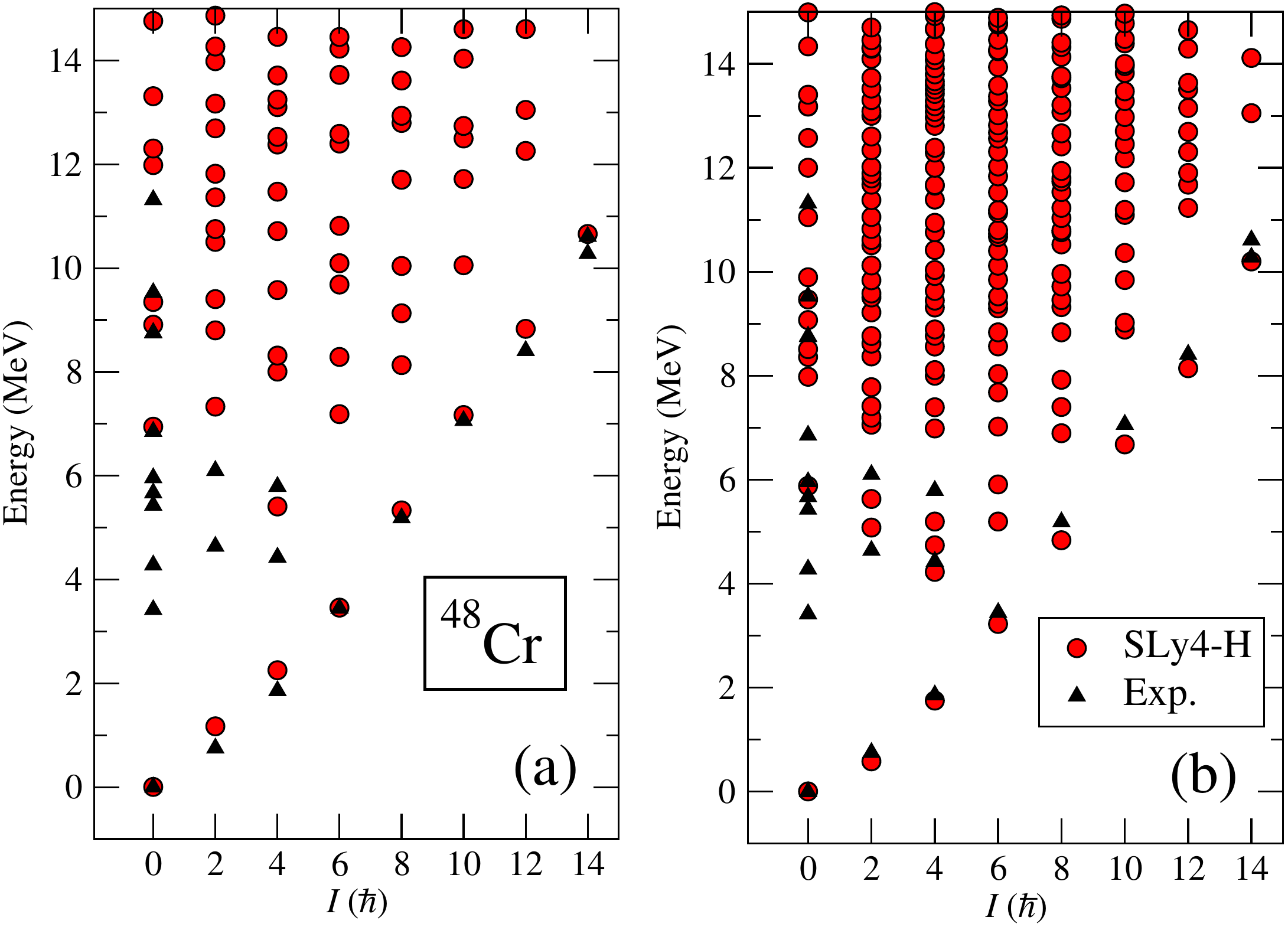}

\caption{Calculated positive parity and even spin spectra for $^{48}$Cr, (a)
without temperature excitations and in panel (b) with inclusion of
single-particle type excitations using the temperature method (\ref{excop})
with $b=0.45$. In both panels we used the SLy4-H Hamiltonian with
11 oscillator shells ($N_{max}=10$) for the single-particle basis
and a grid consisting of $N_{\phi}=191$ HFB vacua for the many-body
basis. Experimental data taken from \citep{Experiment}. The Hamiltonian
for each spin has been diagonalized using the number of natural states
found from the Hill-Wheeler equation (\ref{eq:HW}) considering the
yrast state. \label{fig:temp}}
\end{figure}

The operator acts separately on neutron and proton parts of the states
with the result that  the lowest two-quasiparticle excitations are
added to the states with weights determined by the parameter $b$.
Multi-quasiparticle excitations will also be added due to the structure
of the series but with diminishing weights. The random sign ensures
that even if the energy surface is over sampled the states will still
have orthogonal components allowing for the extraction of more independent
solutions. The quasiparticles are taken to have preserved parity and
signature ($r_{x}\phi$=$e^{-i\pi\hat{j}_{x}}\phi$) quantum numbers.
The quasiparticle pairs in Eq.~(\ref{eq:qp-pairs}) are restricted
to belong to the group with positive parity and signature $\left(\pi,r_{x}\right)=\left(1,1\right)$
so that when acting on an HFB state, the generated excitations do
not change the symmetry of the state. That is, the matrix elements
$z_{k,k'}$ can be characterized by the quantum numbers of the quasiparticle
pairs and these are restricted to have positive signature and parity
and to be of the same nucleon species. After acting on one of the
basis states, the new state obtained contains a mixture of multi-quasiparticle
excitations with random signs that reduces overcompleteness of the
basis. In this way, this temperature inspired method introduces particle
type excitations into the basis in order to complement the more collective
excitations introduced through the generating coordinates. 

An example is shown in Fig.~\ref{fig:temp}. As seen from this figure
the application of the excitation operator allows for a much larger
number of eigenstates to be found and an improved convergence of the
yrast band.

\subsection{Efficient symmetry restoration}

After generating the many-body basis, the following step consists
in computing the matrix elements for the projected overlap and Hamilton
operator. 

By construction, the HFB vacua do not have either a good number of
nucleons or angular momentum. As a consequence, it is necessary to
introduce projection operators to restore symmetries and evaluate
states with definite values for quantum numbers. The matrix elements
read,

\begin{align}
O_{IK,JK'} & =\left\langle \phi_{I}\left|\hat{P}_{MK}^{I\dagger}\hat{P}_{MK'}^{I}\hat{P}_{Z}\hat{P}_{N}\right|\phi_{J}\right\rangle \nonumber \\
 & =\left\langle \phi_{I}\left|\hat{P}_{KK'}^{I}\hat{P}_{Z}\hat{P}_{N}\right|\phi_{J}\right\rangle \label{eq:ov_mat}\\
 & =\sum_{i}w_{i}\left\langle \phi_{I}\left|\hat{R}_{i}\right|\phi_{J}\right\rangle 
\end{align}

\begin{align}
H_{IK,JK'} & =\left\langle \phi_{I}\left|\hat{P}_{MK}^{I\dagger}\hat{H}\hat{P}_{MK'}^{I}\hat{P}_{Z}\hat{P}_{N}\right|\phi_{J}\right\rangle \nonumber \\
 & =\left\langle \phi_{I}\left|\hat{H}\hat{P}_{KK'}^{I}\hat{P}_{Z}\hat{P}_{N}\right|\phi_{J}\right\rangle \nonumber \\
 & =\sum_{i}w_{i}\left\langle \phi_{I}\left|\hat{H}\hat{R}_{i}\right|\phi_{J}\right\rangle \label{eq:ham_mat}
\end{align}
where $\hat{P}$ are projection operators for neutron, proton number,
and angular momentum. $\hat{R}_{i}=\hat{R}\left(\alpha_{i},\beta_{i},\gamma_{i},\theta_{i}^{Z},\theta_{i}^{N}\right)$
are the rotations over the gauge and Euler angles, with the corresponding
weights $w_{i}$ (see e.g. \citep{enami99,Bally_rot}). From each
HFB state one can often project out several many-body states with
different $K$ projections.  The final energies are invariant with
respect to the orientation in the laboratory frame, so these matrix
elements do not depend on the $M$ quantum number.

The computation of the matrix elements above is time consuming due
to the projection operators, which involve angular integrations over
angles in space and gauge space. As a consequence, it is essential
to be able to perform accurate and systematic truncations in the computations
of these matrix elements. Our truncation scheme is based on the Bloch
Messiah decomposition, which allows to rewrite the Bogoliubov matrices
$U$ and $V$ as $U=D\bar{U}C$, $V=D^{*}\bar{V}C$ where $D$ and
$C$ are both unitary matrices and $D$ defines the so-called canonical
basis associated with the Bogoliubov vacuum \citep{rin80}. $\bar{U}$
and $\bar{V}$ can be chosen as diagonal and skew-symmetric, respectively.
The matrix $\bar{V}$ is written in terms of blocks of dimension $2\times2$
with elements ($v_{i}$, -$v_{i}$) where $v_{i}^{2}$ is the occupation
probability of the canonical basis state $i$ (the matrix elements
$u_{i}$ of $\bar{U}$ are such that $u_{i}^{2}+v_{i}^{2}=1$). Our
truncation criteria is defined by first sorting the $N$ occupation
numbers $v_{i}^{2}$ in descending order (see \citep{car21}) and
then truncating the canonical basis, that is we consider a smaller
size $n\leq N$ where $n$ is such that 
\begin{equation}
\sum_{i}v_{i}^{2}-\sum_{i}^{n}v_{i}^{2}<0.01.\label{eq:n-dim}
\end{equation}
The occupation numbers differ for each state so each state is thus
truncated differently and stored in the smaller representation before
calculation of the matrix elements. The truncated states thus define
our new basis states where long tails and numerical noise have been
removed. Using these truncated states, the overlaps of rotated Bogoliubov
states are computed and all the calculations are reduced to the minimal
occupied subspace using the Bloch Messiah transformation (see \citep{car21,rin80,Yao2009}
and Appendix \ref{sec:Appendix:Computation-of-Matrix}). 

An unpaired HFB vacua will have a dimension $n$ corresponding to
the number of particles and exact zeros outside of this space. Paired
vacua will have varying dimension depending on the pairing distribution.
It is thus essential to be able to calculate overlaps of states having
very different sizes. While the applied overlap formula allows the
calculations of overlaps when states have $v_{i}$'s exactly equal
to zero it also allows to reduce the dimension by keeping only non-zero
$v_{i}$'s thus keeping only the essential information and therefore
greatly reducing the computational time \citep{car21}. 

\subsection{Odd numbers of nucleons\label{subsec:odd-systems}}

The computation for nuclei with an odd number of nucleons proceeds
similarly to the even-even case. In this paper, we consider the case
of even-odd nuclei. Having generated the set of HFB vacua as described
in the previous section, the odd-state basis is generated as a set
of one quasiparticle creation operators acting on the HFB states,
namely: 
\begin{equation}
|\Phi^{odd}\rangle\equiv\beta_{a}^{\dagger}|\Phi^{even}\rangle.\label{eq:odd_basis}
\end{equation}
The corresponding Bogoliubov matrices $U^{odd}$ and $V^{odd}$ are
easily obtained by replacing the $a^{th}$ column in $U$ and $V$
by the corresponding column in $V^{*}$, $U^{*}$ \citet{rin80}.
Obviously, as in the case of even nuclei, the ability to truncate
in a systematic manner is also critical for an efficient computation
in the odd case. In this context, the computation of the overlap for
odd system is performed using the truncated formula in \citet{car21}.
The application of the Bloch Messiah decomposition allows to rewrite
the odd vacua as a product of three matrices 
\begin{eqnarray}
U^{odd} & = & D^{odd}\bar{U}^{odd}C^{odd}\;,\\
V^{odd} & = & D^{odd*}\bar{V}^{odd}C^{odd}\,,
\end{eqnarray}
where $D^{odd}$ and $C^{odd}$ are unitary matrices and,
\begin{equation}
{\bar{U}}^{odd}=\begin{pmatrix}0 & 0 & 0 & 0 & 0\\
0 & 1 & 0 & 0 & 0\\
0 & 0 & u_{2} & 0 & 0\\
0 & 0 & 0 & u_{2} & 0\\
0 & 0 & 0 & 0 & \ddots
\end{pmatrix},{\bar{V}}^{odd}=\begin{pmatrix}1 & 0 & 0 & 0 & 0\\
0 & 0 & 0 & 0 & 0\\
0 & 0 & 0 & v_{2} & 0\\
0 & 0 & -v_{2} & 0 & 0\\
0 & 0 & 0 & 0 & \ddots
\end{pmatrix}.
\end{equation}
The structure of $\bar{U^{odd}}$ and $\bar{V^{odd}}$ is almost identical
to the even-even case except for the ``odd'' particle, which is
unpaired. This unpaired particle is, by convention, placed in the
first position in both matrices. The truncations of matrices can then
proceed similarly as in the even case. That is, the truncation is
dictated by the values $v_{i}$ of the paired particles. After decomposing
the matrices in this form the calculation of projected Hamiltonian
matrix elements proceeds similarly as in the even case (see appendix
\ref{sec:Appendix:Computation-of-Matrix}).

\subsection{Hill-Wheeler equation}

The spectra are obtained by solving the Hill-Wheeler (HW) equation
\citep{rin80}. The state solutions of the HW equations are, by construction,
eigenstates of the parity and angular momentum operators $I^{2}$
and $I_{z}$.

In matrix form the Hill-Wheeler equation reads,

\begin{equation}
Hh=EOh,\label{eq:HW}
\end{equation}
with $H$ from Eq.~(\ref{eq:ham_mat}), $O$ from Eq.~(\ref{eq:ov_mat})
and where $h$ and $E$ are the resulting eigenvector and eigenvalue
solutions. This equation can be solved separately for each total angular
momentum $I$ giving energies $E_{n}^{I}$ and corresponding eigenstates
as expansions in terms of the projected HFB states:

\begin{equation}
\ket{IM,n}=\sum_{a=1}^{N_{a}}\sum_{K=-I}^{I}h_{aK,n}^{I}\hat{P}_{MK}^{I}\hat{P}^{N}\hat{P}^{Z}\ket{\phi_{a}}.\label{eq:|IM>}
\end{equation}
In this equation, $\ket{\phi_{a}}$ are the $N_{a}$ HFB-basis states
and the operators $\hat{P}$ are projection operators for proton number,
neutron number and angular momentum. The coefficients $h_{aK,n}^{I}$
are found from the Hill-Wheeler equation in the basis of projected
HFB-states and scaled such that $\ket{IM,n}$ becomes normalized.

\subsection{Transitions and Quadrupole Moments}

Because angular momentum projection is performed, the model gives
eigenstates in the laboratory system as output. This makes it natural
and straightforward to calculate  observables  avoiding the process
of extracting them from the internal system; which inevitably contains
approximations. 

 Furthermore, since the model allows for calculations in large model
spaces there is no need for effective charges.  

\subsubsection{Reduced transition probability}

Because of the interaction between the charged nucleus and the electromagnetic
field, it is possible to have transitions between eigenstates of the
nuclear Hamiltonian by emitting (or absorbing) a photon. Those transitions
can be classified into electromagnetic multipoles. For a given multipole
of order $\lambda$ the emitted (absorbed) photon will carry a total
angular momentum of $\lambda\hbar$.

The transition rate $T$ (the life time is given by $\tau=\hbar/T$)
from an initial to a final nuclear eigenstate for an electrical multipole
is given by \citep{rin80}, in SI-units
\begin{equation}
T_{fi}^{\lambda\mu}=\dfrac{2}{\varepsilon_{0}\hbar}\dfrac{\lambda+1}{\lambda[(2\lambda+1)!!]^{2}}\left(\dfrac{E_{\gamma}}{\hbar c}\right)^{2\lambda+1}\vert\bra{\Psi_{f}}\hat{Q}_{\lambda\mu}\ket{\Psi_{i}}\vert^{2},
\end{equation}
where $E_{\gamma}$ is the energy of the emitted photon. This expression
is derived from "Fermi's golden rule" up to first order in perturbation
theory.

Due to the fact that quadrupole deformations are the dominant shape
degrees of freedom for atomic nuclei, the quadrupole mode is the most
prominent one for the radiation. Hence, here we will consider $E2$-transitions.

A nuclear eigenstate has definite values for the total angular momentum
$I$ and its projection $M$. Often one do not want to distinguish
between different $M$-values; neither for final nor initial states.
Therefore one averages over initial $M$ (assuming an equal distribution
of initial $M$-values) and sum over final $M$ (the final $M$-value
is not important). This type of rate is therefore given by (for $\lambda=2$)
\begin{equation}
\begin{aligned}T_{fi}^{\lambda=2} & =\dfrac{1}{2I_{i}+1}\sum_{M_{f}M_{i}\mu}T_{fi}^{2\mu}\\
 & =\dfrac{1}{75\varepsilon_{0}\hbar}\left(\dfrac{E_{\gamma}}{\hbar c}\right)^{5}\dfrac{1}{2I_{i}+1}\sum_{M_{f}M_{i}\mu}\vert\bra{I_{f}M_{f}}\hat{Q}_{2\mu}\ket{I_{i}M_{i}}\vert^{2}\\
 & \equiv\dfrac{1}{75\varepsilon_{0}\hbar}\left(\dfrac{E_{\gamma}}{\hbar c}\right)^{5}B(E2;I_{i}\rightarrow I_{f}),
\end{aligned}
\label{eqn:B_def}
\end{equation}
where the reduced transition probability $B$ has been defined. The
$B(E2)$-values do not contain the large gamma-ray energy dependence
of the transition rate. Therefore, calculations of the reduced transition
probability are more easily compared to experiment than the transition
rate.

\subsubsection{Projection and GCM}

In the model presented in this paper, projections onto good particle
number and angular momentum are performed. In this approach, the $n$:th
state for given $I$ and $M$ can be written as in Eq.~(\ref{eq:|IM>}).

With those states, the matrix element for the reduced transition probability
becomes, 
\begin{equation}
\begin{aligned}\bra{I^{\prime}M^{\prime},n^{\prime}} & \hat{Q}_{2\mu}\ket{IM,n}=\\
 & =\sum_{a^{\prime},a=1}^{N_{a}}\sum_{K^{\prime}=-I^{\prime}}^{I^{\prime}}\sum_{K=-I}^{I}h_{a^{\prime}K^{\prime},n^{\prime}}^{I'*}h_{aK,n}^{I}\\
 & \times\bra{\phi_{a\prime}}\hat{P}_{M^{\prime}K^{\prime}}^{I^{\prime}\dagger}\hat{Q}_{2\mu}\hat{P}_{MK}^{I}\hat{P}^{N}\hat{P}^{Z}\ket{\phi_{a}},
\end{aligned}
\end{equation}
Where it has been used that $\hat{Q}_{2\mu}$ conserves particle number.
In reference \citet{enami99} it is stated that,
\begin{equation}
\begin{aligned}\hat{P}_{M^{\prime}K^{\prime}}^{I^{\prime}\dagger}\hat{Q}_{2\mu}\hat{P}_{MK}^{I} & =C_{IM2\mu}^{I^{\prime}M^{\prime}}\sum_{\nu}C_{I,K^{\prime}-\nu,2\nu}^{I^{\prime}K^{\prime}}\hat{Q}_{2\nu}\hat{P}_{K^{\prime}-\nu,K}^{I}\end{aligned}
,
\end{equation}
where the $C$:s are Clebsch-Gordan coefficients with notation such
that the two angular momenta in the subscript couple to the angular
momenta in the superscript. Hence, in total we get 
\begin{equation}
\begin{aligned} & \bra{I^{\prime}M^{\prime},n^{\prime}}\hat{Q}_{2\mu}\ket{IM,n}=C_{IM2\mu}^{I^{\prime}M^{\prime}}\\
 & \times\sum_{a^{\prime},a=1}^{N_{a}}\sum_{K^{\prime}=-I^{\prime}}^{I^{\prime}}\sum_{K=-I}^{I}h_{a^{\prime}K^{\prime},n^{\prime}}^{I'*}h_{aK,n}^{I}\\
 & \times\sum_{\nu}C_{I,K^{\prime}-\nu,2\nu}^{I^{\prime}K^{\prime}}\bra{\phi_{a^{\prime}}}\hat{Q}_{2\nu}\hat{P}_{K^{\prime}-\nu,K}^{I}\hat{P}^{N}\hat{P}^{Z}\ket{\phi_{a}}.
\end{aligned}
\end{equation}
In the expression for the $B(E2)$-value, Eq.~(\ref{eqn:B_def}),
the summations over $M^{\prime},M$ and $\mu$ only involves the first
Clebsch-Gordan coefficient in the above matrix element. Using orthogonally
relations for the Clebsch-Gordan coefficients, the sum reduces to
\begin{equation}
\sum_{M^{\prime}M\mu}\vert C_{IM2\mu}^{I^{\prime}M^{\prime}}\vert^{2}=\sum_{M^{\prime}}1=2I^{\prime}+1.
\end{equation}
The final expression for the reduced transition probability is then
\begin{equation}
\begin{aligned} & B(E2;I\rightarrow I^{\prime})\\
 & =\dfrac{2I^{\prime}+1}{2I+1}\left|\sum_{a^{\prime},a=1}^{N_{a}}\sum_{K^{\prime}=-I^{\prime}}^{I^{\prime}}\sum_{K=-I}^{I}h_{a^{\prime}K^{\prime},n^{\prime}}^{I'*}h_{aK,n}^{I}\right.\\
 & \left.\times\sum_{\nu}C_{I,K^{\prime}-\nu,2\nu}^{I^{\prime}K^{\prime}}\bra{\phi_{a^{\prime}}}\hat{Q}_{2\nu}\hat{P}_{K^{\prime}-\nu,K}^{I}\hat{P}^{N}\hat{P}^{Z}\ket{\phi_{a}}\right|^{2}.
\end{aligned}
\end{equation}
In the cases where measurements exists, we also compare the spectroscopic
quadrupole moment, defined as:

\begin{equation}
Q_{spec}=\sqrt{\dfrac{16\pi}{5}}\bra{IM=I}\hat{Q}_{20}\ket{IM=I}.\label{eq:Qspec}
\end{equation}
This $Q_{spec}$ is defined in the laboratory frame and becomes identically
zero for $I<1$. 

\section{Results\label{subsec:Multiplets-in-spherical-1}}

To test the developed model, calculations have been performed for
five nuclei: Four even-even nuclei, the three chromium isotopes $^{48,50,52}$Cr,
$^{24}$Mg and the even-odd $^{49}$Cr. The formalism of extracting
the $B(E2)$-transitions and the quadrupole moment has been implemented
only for even-even nuclei. The results are compared both with experiment
and with other theoretical calculations.

The chromium isotopes have been chosen in order to test the model
when going from the deformed $^{48}$Cr to the more spherical $^{52}$Cr.

The experimental values for energies and transitions are taken from
\citep{Experiment} if not otherwise stated. Experimental values for
the spectroscopic quadrupole moments are rare but the few found are
from \citep{Exp_Q}. Some of the transition rates have not been explicitly
given in ref. \citep{Experiment}, but are instead extracted from
gamma energies and lifetimes according to Eq.~(\ref{eqn:B_def}).
In the case where the state decays in several channels, the lifetime
for channel $a$, $\tau_{a}$, can be calculated from the gamma intensity,
$I(\gamma)$, with the expression
\begin{equation}
\tau_{a}=\tau\frac{\sum_{i}I_{i}(\gamma)}{I_{a}(\gamma)},
\end{equation}
where the sum goes over all channels.

\subsection{Generation of collective subspace for the different nuclei\label{subsec:Generation-of-collective}}

The points in the grid are defined by starting at spherical shape.
For each new point the $\gamma$ angle is increased with the golden
angle $\theta\simeq137.508$. The radial distance $\beta$ is increased
with the square root of the number of points. This generates a rather
homogeneously sampled circular area in the $(\beta,\gamma)$ plane.
With cranking included, the surface will have mirror symmetry in the
$y-$axis. An example for $^{48}$Cr with $j_{x}=0$ is shown in 
\begin{figure}
\includegraphics[clip,width=1\columnwidth]{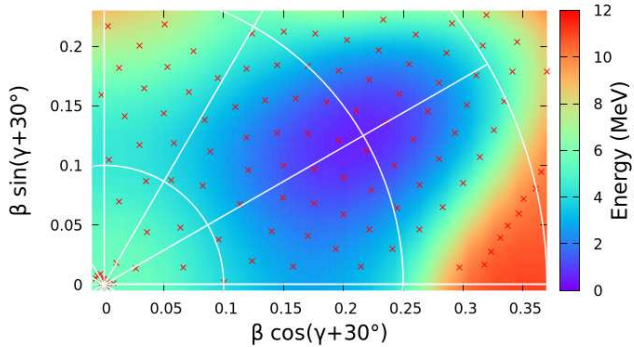}

\caption{HFB energy versus deformation for $^{48}$Cr with the SLy4-H Hamiltonian.
Twelve oscillator shells are used for the single-particle basis ($N_{max}=11$),
$j_{x}=0$ and the pairing interaction strenght is kept fixed $\left(g_{p},g_{n}\right)=\left(1,1\right)$.
The $\left(\beta,\gamma\right)$ plane is drawn with the standard
Lund convention with $\gamma=60^{\circ}$ along the positive $y$-axis.\label{fig:surface}}
\end{figure}
Fig.~\ref{fig:surface}. As seen from this figure $^{48}$Cr has
a prolate minimum centered around $\gamma=0^{\circ}$ with $\beta\sim0.25$.
As $j_{x}$ is increased this minimum will move towards $\gamma=60^{\circ}$
where the rotation eventually becomes non-collective since the shape
is then rotationally symmetric around the cranking axis ($x$-axis). 

For the pairing we use a grid in two variables $g_{p}$ and $g_{n}$.
These grids are constrained to four values: $\{0.6,1.0,1.4,1.8\}$.
These values specify scaling of the pairing $\Delta$ values. A value
of $1.0$ would imply generating the surface with only the self consistent
pairing. While a value of 1.4 implies generating basis states with
a stronger interaction that gives 1.4 times larger pairing gaps.

For the cranking frequency $\omega$ we choose a grid of three different
$j_{x}$ values $\{0,4,8\}$ and the $\omega$ values needed for each
state to obtain those $j_{x}$ values are estimated as $\omega=\frac{1}{2\mathcal{J}(\beta,\gamma,\Delta)}\left(2j_{x}+1\right)$
\citep{nilsson1995}. The moment of inertia is estimated as in \citep{BENGTSSON1986},
where for simplicity we used $\Delta=0.1$ MeV for all points. For
each point in the $\left(\beta,\gamma\right)$ plane values of $g_{p},g_{n}$
and $j_{x}$ are randomly drawn from the allowed sets in order to
create states that sample the relevant many-body space. For the even-even
nuclei we have used basis states with signature $r_{x}=1$ and for
the odd $^{49}$Cr we compare the use of both $r_{x}=\pm i$.

In the numerical calculation of the excitation operator (Eq.~(\ref{excop}))
we have used $b=0.45$ ( see Eq.~(\ref{kt})). This implies that
for the particle species that is easiest to excite the lowest two-quasiparticle
excitation within the considered symmetry group is added to the state
with a weight of 0.45.

For all even-even chromium isotopes, the same calculation parameters
have been used (values for $^{49}$Cr and $^{24}$Mg are given below).
That is, 12 major shells in the harmonic oscillator basis generated
by an updated version of the code \pr{HOSPHE} \citet{CARLSSON20101641}.
The $(\beta,\gamma)$-plane has been sampled with 300 states within
$\beta\leq0.5$ and $-30{^\circ}\leq\gamma\leq150^{{^\circ}}$. Since
we are interested in the low-energy part of the spectra it is sufficient
to consider basis states up to a given cutoff. Therefore only states
within 12 MeV from the state with the lowest calculated energy has
been kept, which is sufficient to cover the energy range of the experimental
yrast states. This resulted in basis sizes of 198, 217 and 206 for
the $^{48}$Cr ,$^{50}$Cr and $^{52}$Cr isotopes, respectively.
For the numerical computation of the projections, 10 points have been
used in the number projections for both types of nucleons and (9,
18, 36) points in the $(\alpha,\beta,\gamma)$ angles for the angular
momentum projection. These number of points for the angular momentum
projection are obtained after applying symmetries to reduce the integration
interval and corresponds to (36, 36, 36) points in the full space,
see e.g. \citep{enami99,Bally_rot}.

The $B(E2)$-values are calculated for every transition that differ
with 2 or 0 units of spin. However, with few exceptions discussed,
only transitions over 10 W.u are plotted in the spectra.

\subsection{$^{48}$Cr}

\begin{figure}
\includegraphics[clip,width=0.9\columnwidth]{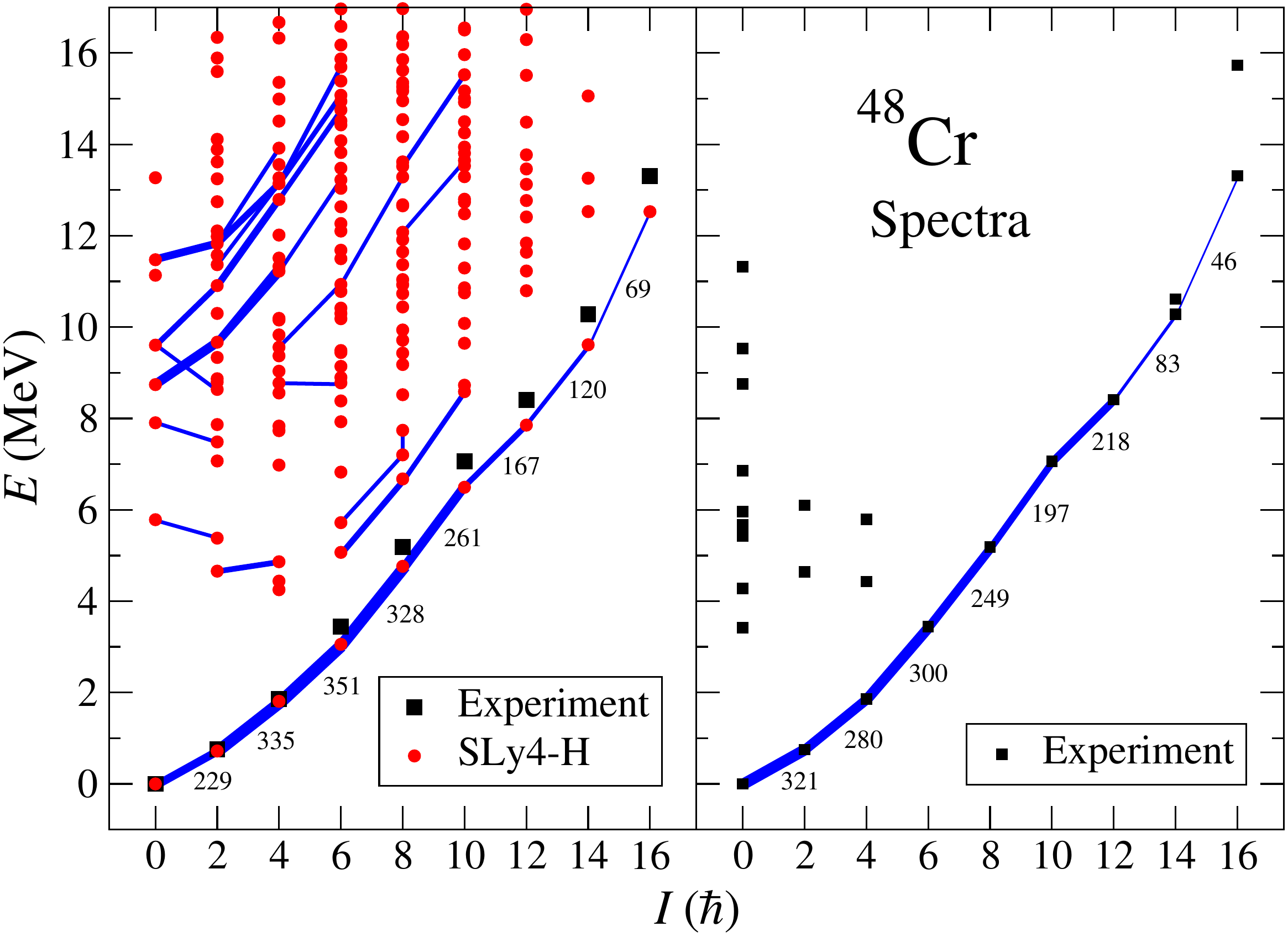}

\caption{Positive parity and even spin spectra including $B(E2)$-transitions
for $^{48}$Cr. Theory to the left and experiment to the right. The
experimental values for the yrast band are also shown together with
theory for easier comparison. The strengths for the transitions are
indicated by the thickness of the lines and are explicitly given in
e$^{2}$fm$^{4}$ for the yrast band.\label{fig:Cr-48_spectra}}
\end{figure}

The spectra for $_{24}^{48}$Cr$_{24}$, both the calculated and the
experimental values, together with the strongest transitions, are
shown in Fig.~\ref{fig:Cr-48_spectra}. The calculation shows the
characteristic of a rotor, $E_{rot}\propto I(I+1)$, for the yrast
band up to spin $10\hbar$ where the first backbending happens. 

The general behavior of a deformed and rotating nucleus approaching
a terminating states has been discussed in reference \citep{AFANASJEV19991}.
For a low angular momentum the rotation is of a collective nature
with the rotation axis perpendicular to the symmetry axis of the nucleus.
With increasing angular momentum the valence nucleons tend to align
their spins with the rotation axis. This continues until all valence
nucleons are fully aligned with the rotation axis. Then no further
angular momentum can be built and one has reached the terminating
state. In that state the rotation axis is parallel with the symmetry
axis and therefore the rotation is of a single particle nature.

In $^{48}$Cr, the termination is expected to happen at $I=16$. This
can be understood from the fact that $^{48}$Cr has four protons and
four neutrons in the $f_{7/2}$ subshell. Aligning all valence nucleons
within this subshell results in spin $16\hbar$. 

The evolution of the internal structure of $^{48}$Cr with angular
momentum up to its terminating state has been investigated in reference
\citep{Juodagalvis_2000} within the cranked Nilsson-Strutinsky (CNS)
model. The conclusion in that paper is that the intrinsic deformation
goes from axially symmetric prolate over triaxial shapes to end up
in a slightly oblate shape when the terminating state is reached.

\begin{figure}
\includegraphics[clip,width=0.9\columnwidth]{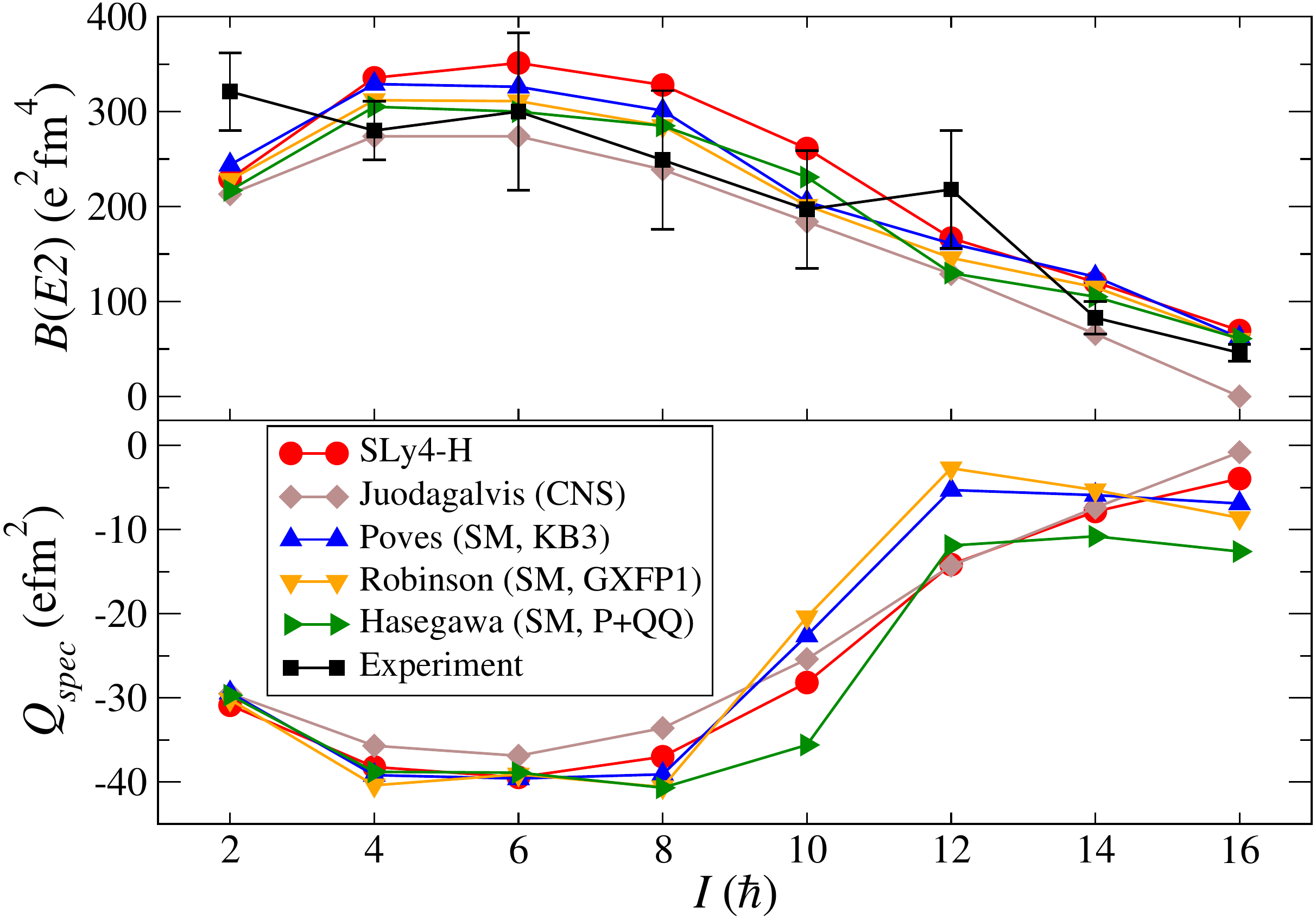}

\caption{$B(E2)$-transitions ($I\rightarrow I-2$) and spectroscopic quadrupole
moments for the yrast band in $^{48}$Cr. No experimental data has
been found for $Q_{spec}$.\label{fig:Cr-48_trans}}
\end{figure}

Comparisons of transitions and quadrupole moments between results
from our model and results obtained from CNS is shown in Fig.~\ref{fig:Cr-48_trans}.
Also three shell model (SM) calculations for three different interactions
in the full $fp$-space are included \citep{Poves_1999,Robinson_2014,HASEGAWA2000_411}.

The results from our Hamiltonian, denoted SLy4-H, are in agreement
with the previous CNS and SM calculations. Both the $B(E2)$- and
the $Q_{spec}$-curves resembles the ones for a rigid rotor up spin
$6\hbar$. For higher angular momentum, the $B(E2)$-values are approaching
zero; showing that the states are indeed of a single particle nature
rather than a mixed collective one. There is also a good agreement
for the prediction of the properties of the spectroscopic quadrupole
moment (\ref{eq:Qspec}). All the models predict a quite drastic change
in the shape associated with the backbend at $I=10-12$ and that the
nucleus becomes close to spherical at the terminating state.

In a $N=Z$ nucleus, as for $^{48}$Cr, it is expected that neutron-proton
pairing should play an important role. This is because both the protons
and the neutrons occupy the same valence space. Thus, they have maximal
spatial overlap, cf. reference \citep{Goodman_1999} for a discussion
of neutron-proton pairing, the different types from the different
isospin channels and their possible effects. Even though our model
does not include neutron-proton pairing, it is interesting to compare
the backbending from the calculations with experiment.
\begin{figure}
\includegraphics[clip,width=0.9\columnwidth]{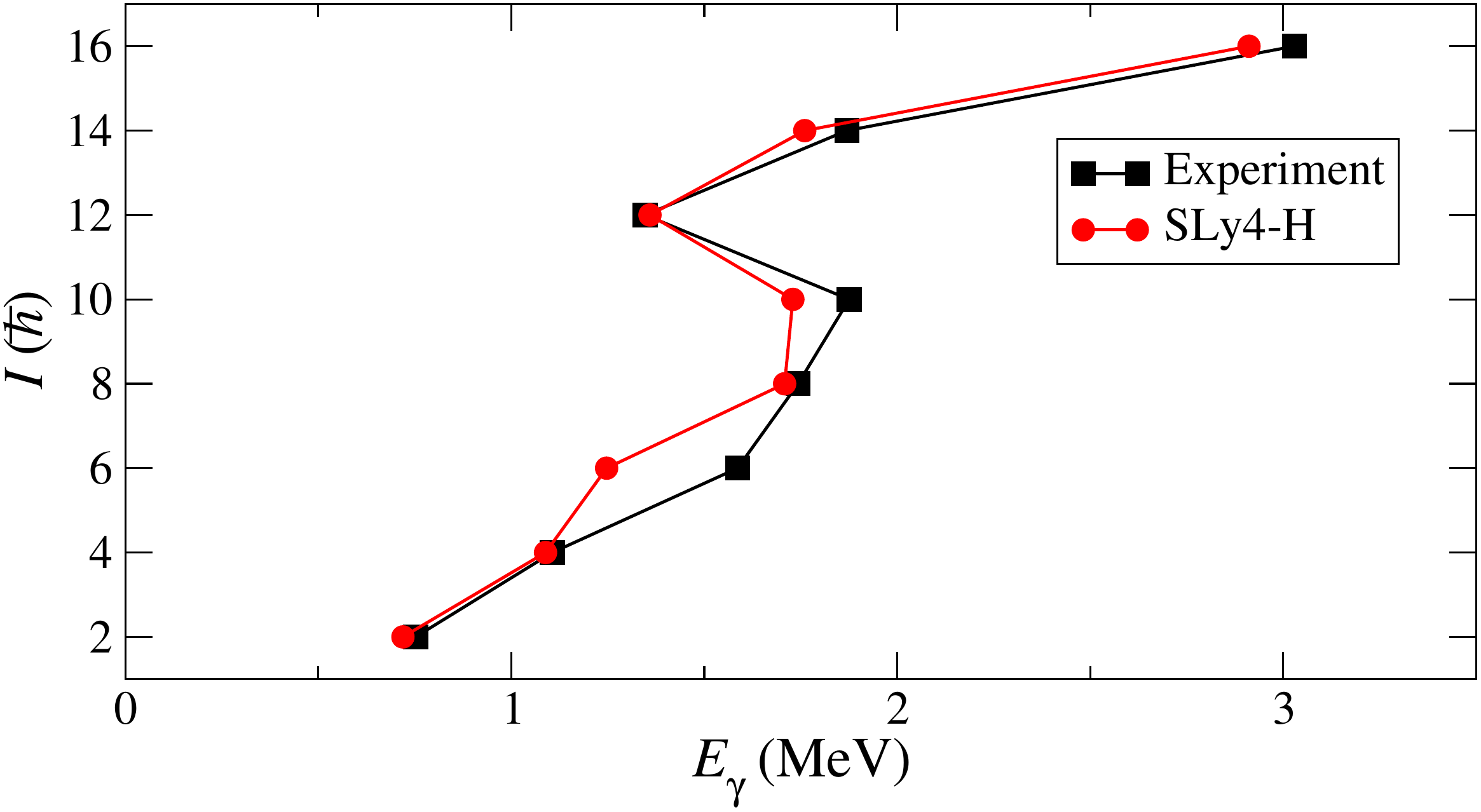}\caption{Released gamma ray energies for transitions $I+2\rightarrow I$ for
the yrast band of $^{48}$Cr. The backbending at $I=10\hbar$ indicates
a change in the internal structure.\label{fig:Cr-48_backbending-1}}
\end{figure}
It can be seen in Fig.~\ref{fig:Cr-48_backbending-1} that the backbending
in SLy4-H occurs at the same spin and have the same magnitude as in
experiment. However, the gamma energies are consistently lower for
all angular momenta (except at the backbending). This is in line with
references \citep{Poves_1999,Robinson_2005}. In both those papers
they separate out the neutron-proton pairing in the isospin $T=0$
channel from the rest of the pairing to see its effect. They both
find that, without this pairing, the spectra becomes suppressed. This
would imply a shift to the left for the backbending curve in Fig.~\ref{fig:Cr-48_backbending-1}
with around 0.2 - 0.5 MeV.

\subsection{$^{49}$Cr}

\begin{figure}
\includegraphics[clip,width=1\columnwidth]{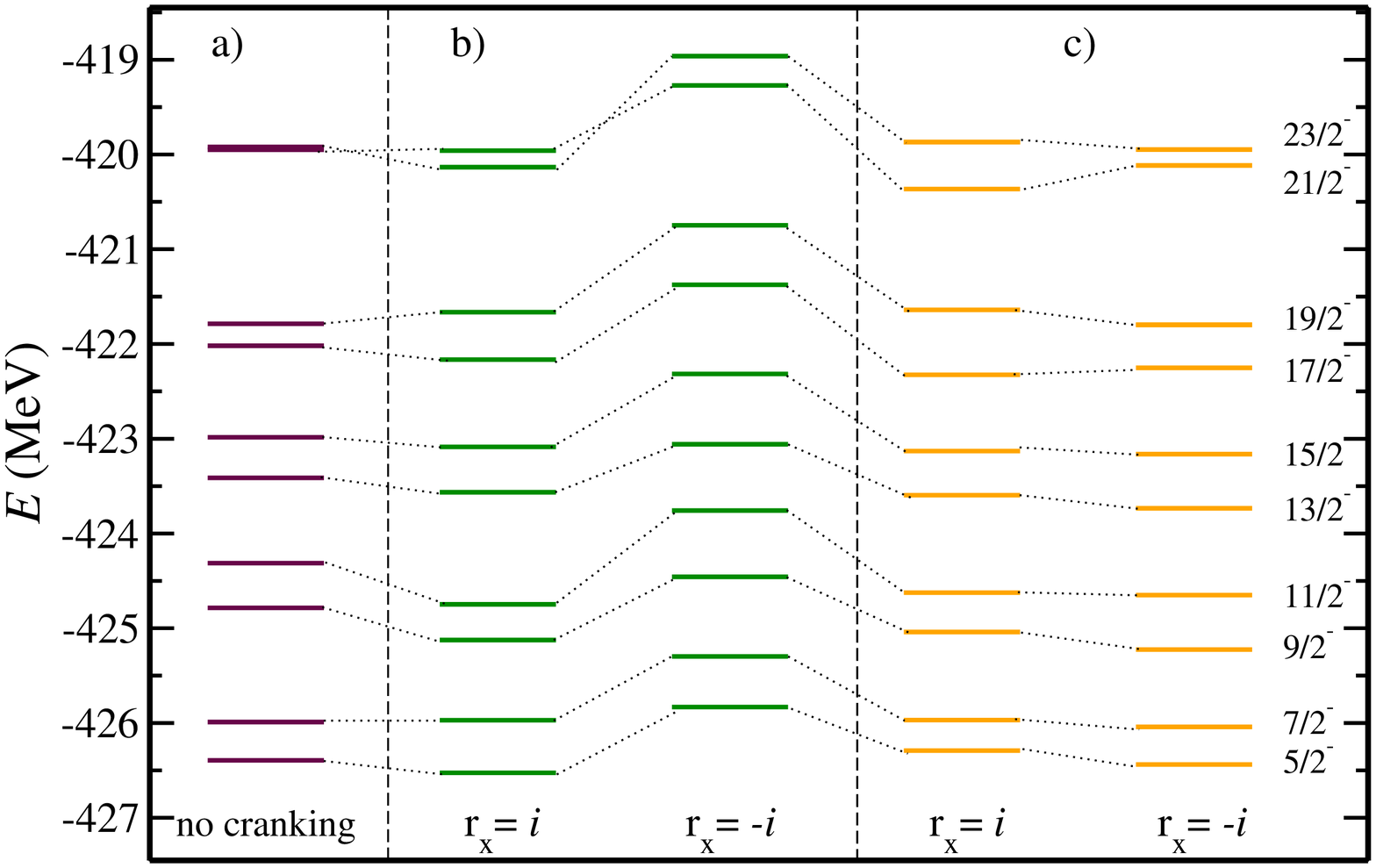} \caption{Energies of the g.s. rotational band in $^{49}$Cr computed in different
bases. In the panel a) the basis include 116 states and the cranking
is switched off. In the panels b) and c), respectively, 114 and 156
states are included in the basis and the cranking is switched on.
Results on the left (right) side of panels b) and c) are obtained
in a basis with signature $r_{x}=i$ ($-i$). \label{all_odd-2-1}}
\end{figure}

We now focus on the even-odd $_{24}^{49}$Cr$_{25}$ isotope. Shell
model calculations in the $pf-$ shell have shown a good reproduction
of the g.s. band and its rotational patterns at low spin can be described
by the Particle-Rotor model as a $K^{\pi}=5/2^{-}$ band based on
the $\nu${[}312{]}5/2- Nilsson orbital \citep{pinedo}. 
Figure \ref{all_odd-2-1} shows the computed energies of states in
the g.s. band of $^{49}$Cr in several bases. The results in panel
a) of Fig.~\ref{all_odd-2-1} are obtained, without cranking, in
a basis of 116 states whereas for the results in panel b) and c),
the cranking is included and the number of basis states is 114 and
156, respectively. Without cranking, the states in the odd basis with
opposite signature are related by time-reversal and consequently the
computed energies are identical whether the basis states have a signature
$r_{x}=i$ or $r_{x}=-i$. This is not the case anymore when the cranking
is included and we show in the panels b) and c) of Fig.~\ref{all_odd-2-1}
results for both signatures. All parameters are the same as for $^{48}$Cr
and for each calculation, the basis is formed by blocking the qp.
(see Eq.~(\ref{eq:odd_basis})) with the lowest energy \footnote{Due to the application of the excitation operator (\ref{excop}) on
the HFB vacua, the qp.'s are no longer eigenstates for a finite value
of $kT$. Nevertheless, the qp. with the lowest energy computed before
the application of the operator (\ref{excop}), is selected to construct
the odd state basis. This is justified by the fact that for low $kT$
value, the ordering of the average qp. energy is not dramatically
affected.}. 

A general property is that states with $I=1/2,5/2,9/2...$ are best
described within a basis with signature $r_{x}=-i$ and states with
$I=3/2,7/2,11/2...$ are best described within a basis with $r_{x}=i$
\citep{bohr1969nuclear}. However, from Fig.~\ref{all_odd-2-1} one
notices that the difference in the energies computed with the different
signatures diminishes when the basis increases. It is then more natural
to consider as the most precise energy for a state of angular momentum
$I$, the lowest energy among the two energies computed in the largest
basis with $r_{x}=i$ and $r_{x}=-i$. As one can see in Fig.~\ref{all_odd-2-1},
the difference in the g.s. energy obtained with $r_{x}=i$ and $r_{x}=-i$
is $\sim700$ keV for the basis made of 114 states (panel b) and decrease
to less than 150 keV in the larger basis (panel c). It is also worth
noting that the difference between the lowest computed g.s. energies
with cranking (panels b and c) and the g.s. energy without cranking
(panel a) is $\sim100$ keV. \textbf{}

Taking the lowest state in the bigger basis we obtain a binding energy
$B^{th}=-426.395$ MeV ( that is the energy of the g.s. $I^{\pi}=5/2^{-}$
in panel c) of Fig.~\ref{all_odd-2-1} in the basis with $r_{x}=-i$),
which is slightly lower than the experimental binding energy $B^{exp}=-422.051$
MeV. In order to gain some insights into the amount of correlations
included beyond the mean-field, it is instructive to compare $B^{th}$
with the lowest mean-field energy among the basis states. For each
odd state $|\Phi^{a,i}\rangle\equiv\beta_{a}^{\dagger}|\Phi_{i}\rangle$,
we can assign a mean-field energy $E_{0}^{(a,i)}=E_{0}^{i}+e_{a}$,
with $E_{0}^{i}$ the HFB-energy of the even-even vacuum $|\Phi_{i}\rangle$
and $e_{a}$ the energy of the qp. $a$. In that particular case,
the lowest mean-field energy among the basis states is $-420.560$
MeV, which implies that the beyond mean-field effects included in
the theory, lower the energy by $\sim5.8$ MeV.

\begin{figure}
\includegraphics[clip,width=1\columnwidth]{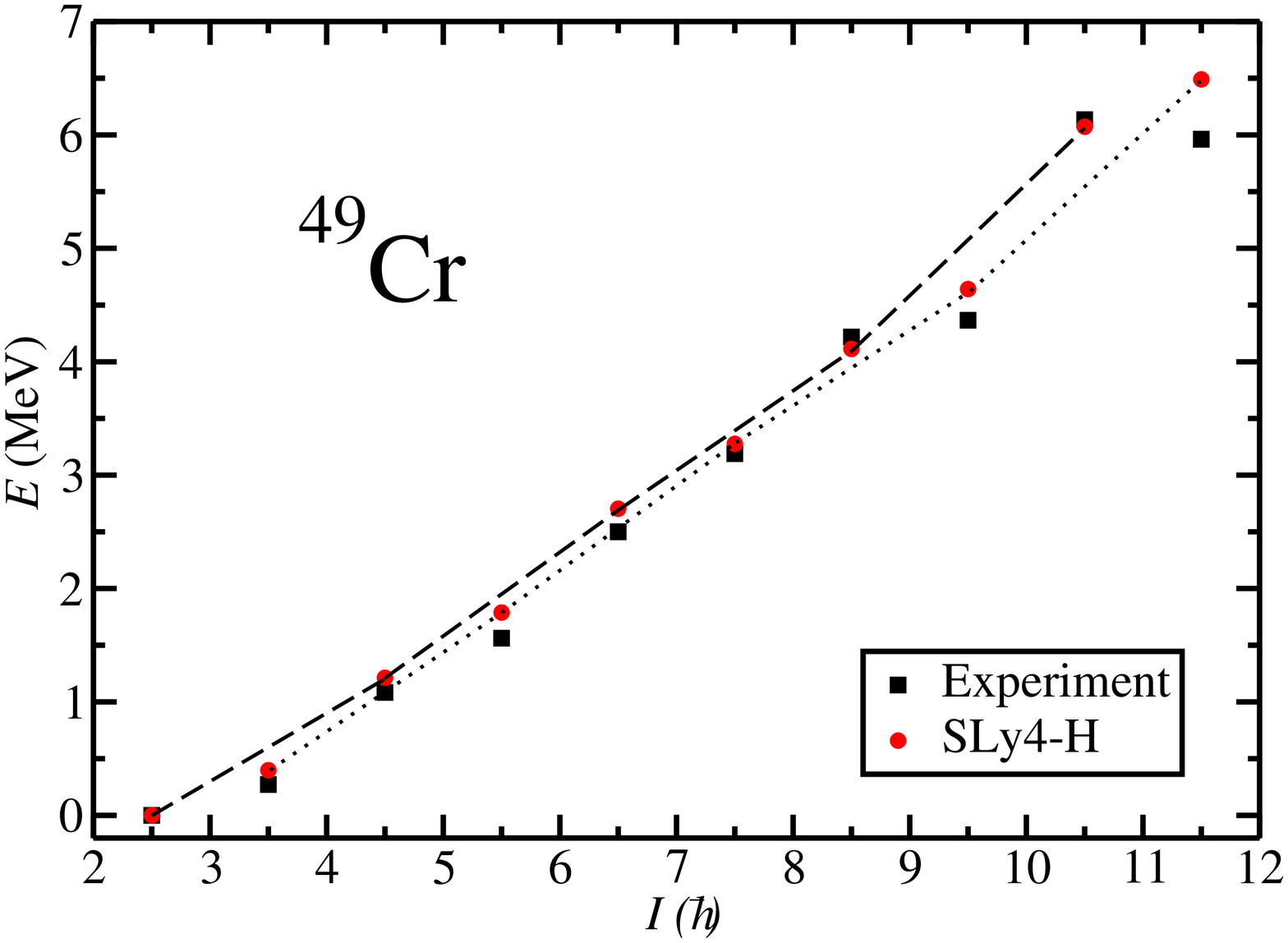} \caption{Excitation energies of states in the g.s. rotational band in $^{49}$Cr.
The black squares denote experimental data, whereas the red circles
show the results of the computation in the largest basis considered
(156 states) with cranking. For each $I$ the computed energies corresponds
to the lowest energies among the spectra displayed in panel c) of
Fig.~\ref{all_odd-2-1}. The dash and dotted lines denotes the two
branches, which connect the calculated states with $\Delta I=2$ (see
text). \label{fig:Cr49-4-1}}
\end{figure}

We show in Fig.~\ref{fig:Cr49-4-1}, a comparison between the computed
excitation energies and the experimental data. As one can see, the
data are well reproduced by the calculation. In particular, our calculations
reproduce the occurrence of a backbending for spin $19/2\hbar$. 
The g.s. rotational band splits into two branches corresponding to
sequences of states $\Delta I=2$. As one can see in Fig.~\ref{fig:Cr49-4-1}
at low spin, the two branches are close to each other and start to
diverge for larger $I$.

\subsection{$^{50}$Cr}

\begin{figure}
\includegraphics[clip,width=0.9\columnwidth]{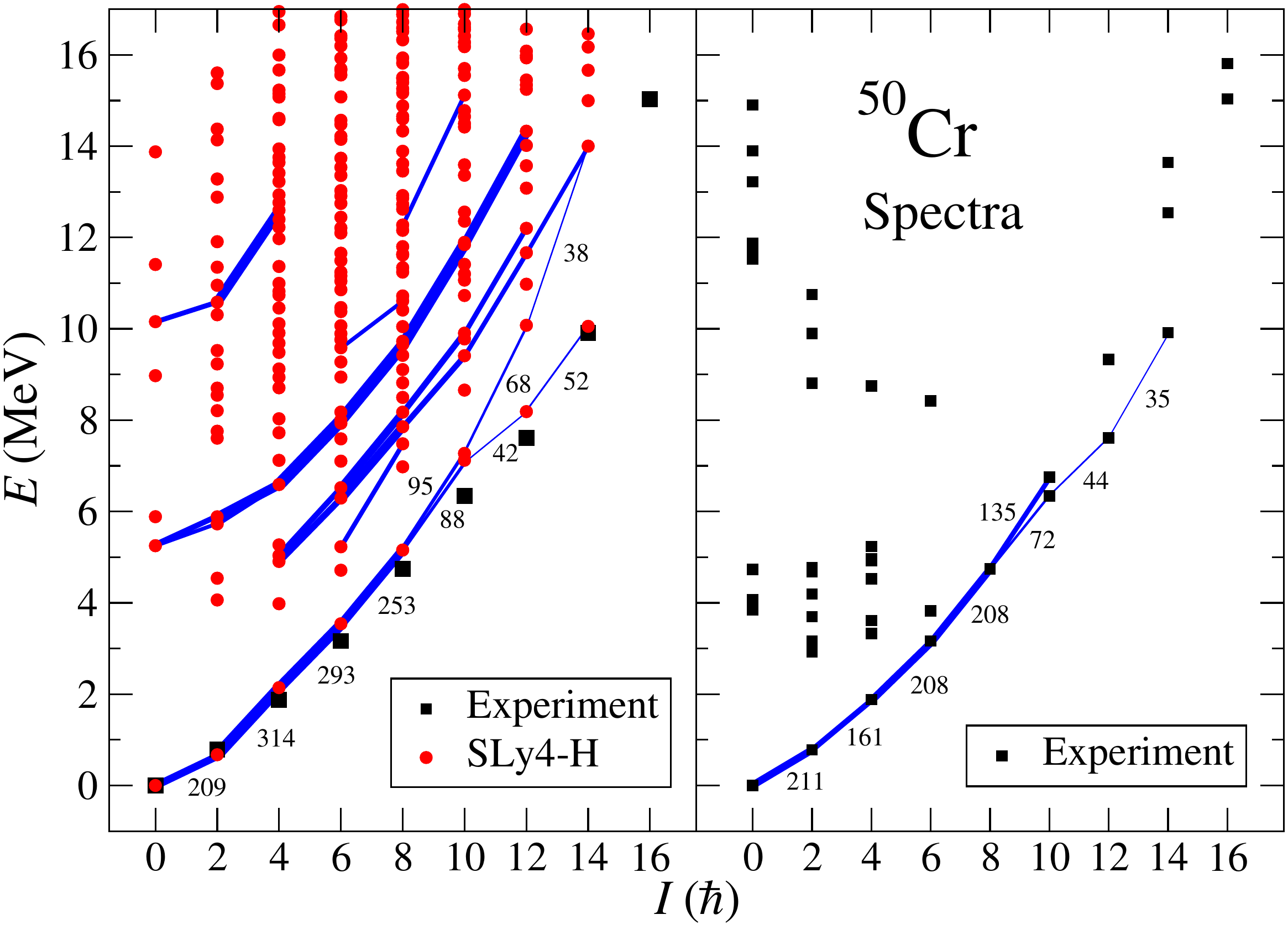}\caption{Same as Fig.~\ref{fig:Cr-48_spectra} for the spectra and $B(E2)$-transitions
in $^{50}$Cr.\label{fig:Cr-50_spectra}}
\end{figure}

The spectrum for $_{24}^{50}$Cr$_{26}$ is given in Fig.~\ref{fig:Cr-50_spectra}.
We calculated up to spin $14\hbar$, which is expected to be the terminating
spin of the ground state configuration.

\begin{figure}
\includegraphics[clip,width=0.9\columnwidth]{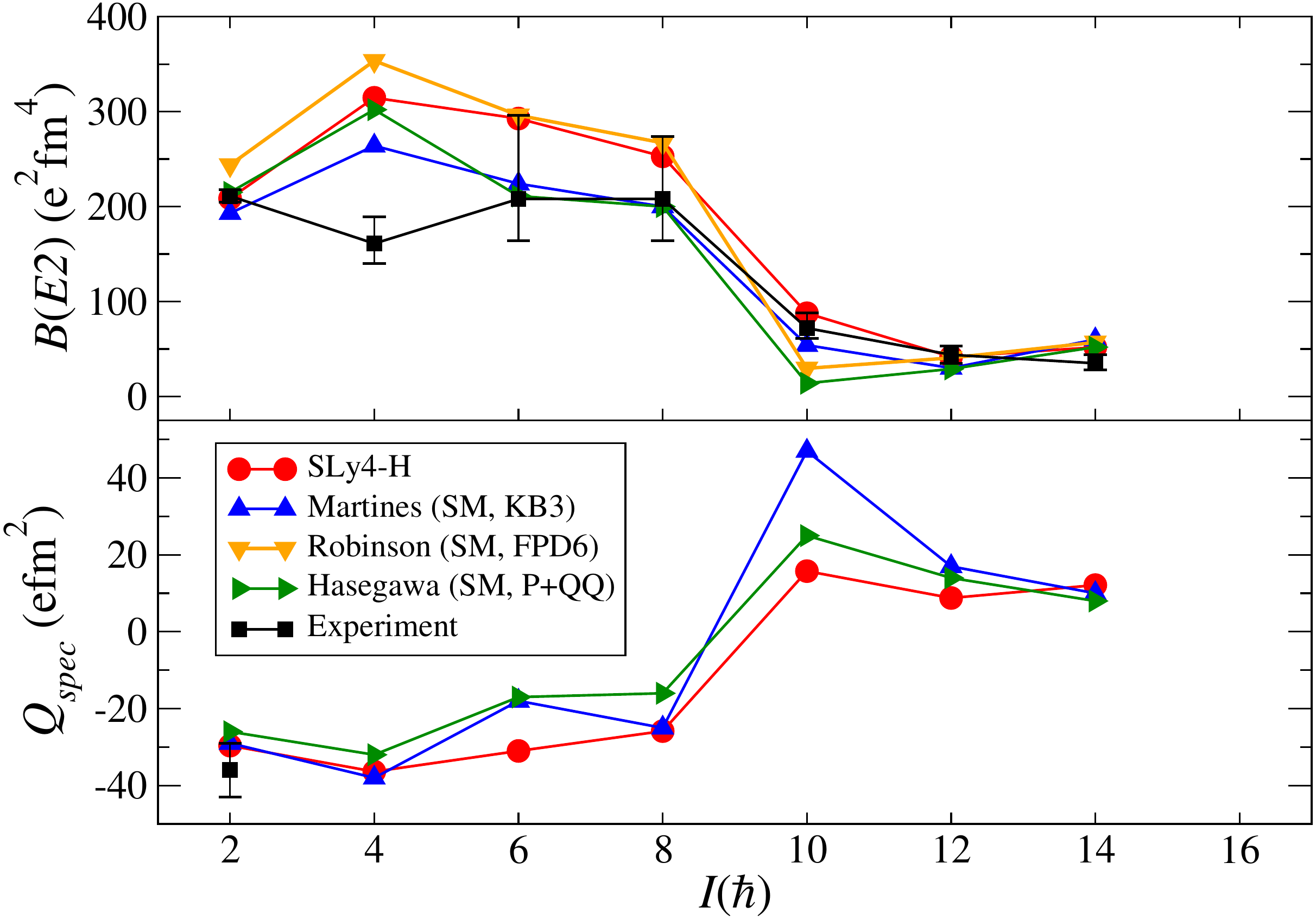}\caption{$B(E2)$-transitions and spectroscopic quadrupole moments for the
yrast band in $^{50}$Cr.\label{fig:Cr-50_B_Q}}
\end{figure}

The calculations reproduce the two lowest $10^{+}$ states that are
very close in energy and that are also seen in experiment. In fact,
the calculations suggest that the ground state band, which starts
from the first $0^{+}$-state, continues up to $I=14$ via the $10_{2}^{+}$
state. The yrast states for $I=10$, $12$ and $14$ seem to originate
from a different band. This is further confirmed from the spectroscopic
quadrupole moments in Fig.~\ref{fig:Cr-50_B_Q}. At $I=10$ they
change sign, indicating a change in the internal structure. 

Experiments also show a stronger $B(E2)$-transition from the $10_{2}^{+}$
state than from the $10_{1}^{+}$ state. Unfortunately, there are
no data of transition strengths for higher spin states above the yrast
band. 

For the purpose of this paper, CNS calculations for $^{50}$Cr have
been performed. The results from those calculations can be used to
interpret the internal structure of the states. It was found that
the $10_{2}^{+}$ state indeed gets its angular momentum from collective
rotation; in the same way as the lower part of the yrast band does.
Whereas the $10_{1}^{+}$ state is predicted to be prolate with the
symmetry axis parallel to the rotational axis. This implies that the
rotation is built up by single particles spins.

Furthermore, the CNS calculations predict two more bands with positive
parity. One band is located around 2.2 MeV above the yrast band and
is built upon a $1p-1h$ excitation of the neutrons. This band is
in fact divided into two nearly degenerated bands with opposite signature.
Another excited band is found around 3.7 MeV above the yrast band
and is built upon a $2p-2h$ neutron excitation. 

As seen in Fig.~\ref{fig:Cr-50_spectra}, our model produce the same
band structure as the CNS calculations. Hence, for this nucleus, the
two methods are consistent with each other.

In experiments no excited bands with positive parity and even spins
have yet been identified.

In Fig.~\ref{fig:Cr-50_B_Q} our results for the transitions and
spectroscopic quadrupole moments are compared with experiment and
shell model calculations for three different interactions \citep{Martinez,Robinson_2005,HASEGAWA2000_411}.

It is interesting to note the discrepancy between experiment and all
of the shown theoretical calculations for the $B(E2;4\rightarrow2)$-value.
Also, the experimental value is not what one would expect from a rotational
model. In fact, the transitions for $^{48}$Cr also do not follow
a rotor description for low angular momenta. But for that nucleus
it is the $B(E2;2\rightarrow0)$ value that is high. So, for both
$^{48}$Cr and $^{50}$Cr the ratio $B(E2;4\rightarrow2)/B(E2;2\rightarrow0)$
is less than 1. This in contrast of the rotational model where the
ratio is 1.43; which is in more agreement with the calculations. This
discrepancy has been pointed out before \citep{B(E2)-ratio}. In reference
\citep{B(E2)-ratio_2} it is suggested the that unusual ratio can
be understood in a collective picture with the inclusion of appropriate
three-body forces. This is shown explicitly for $^{170}$Os in the
interacting boson model.

\subsection{$^{52}$Cr}

\begin{figure}
\includegraphics[clip,width=0.9\columnwidth]{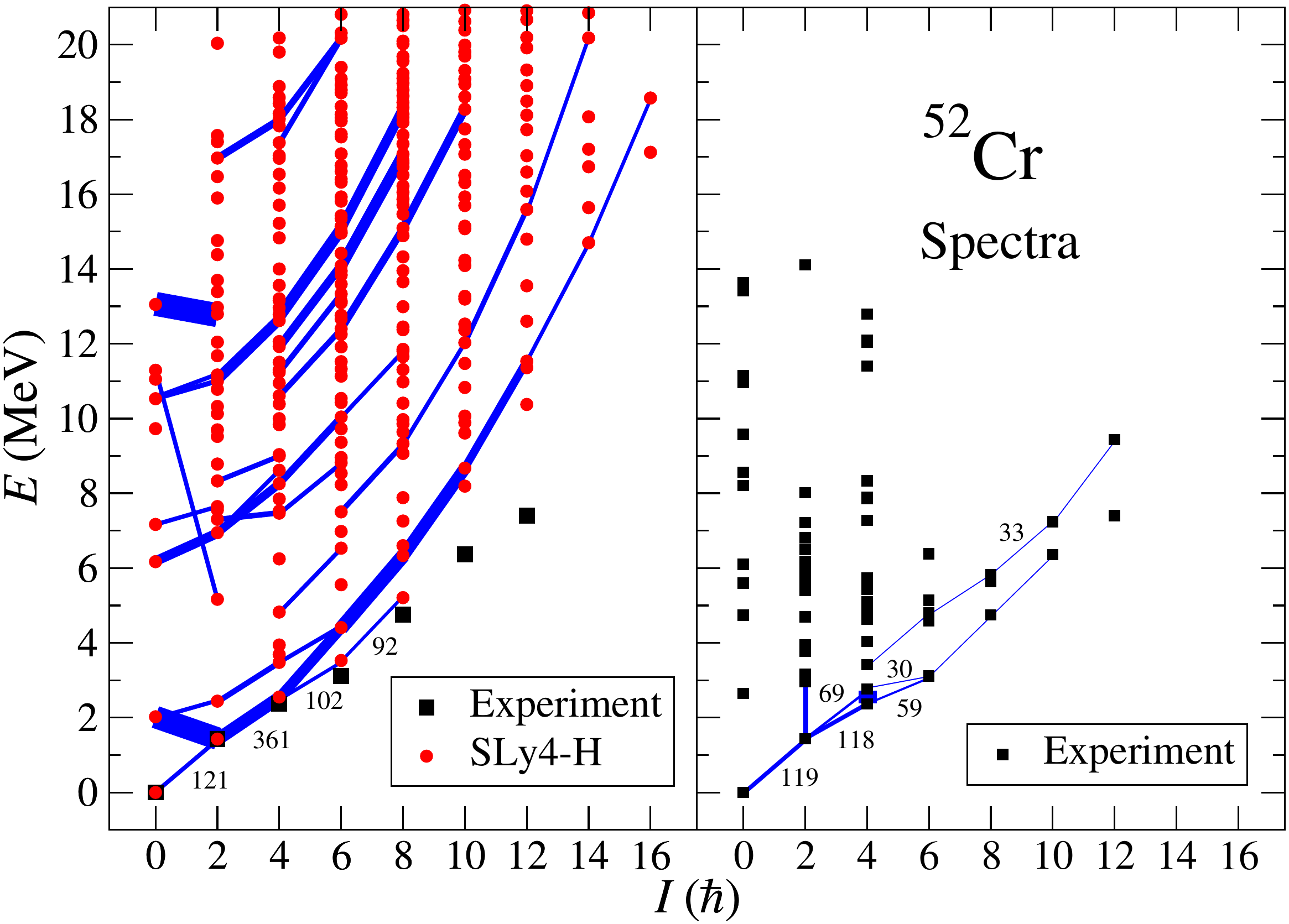}\caption{Same as Fig.~\ref{fig:Cr-48_spectra} for the spectra and $B(E2)$-transitions
in $^{52}$Cr. The two identified experimental bands are indicated
with thin lines when there are no measured $B(E2)$ values. \label{fig:Cr-52_spectra}}
\end{figure}

In $_{24}^{52}$Cr$_{28}$ the eight valence neutrons fill up the
$f_{7/2}$ orbit and therefore the nucleus is expected to be less
deformed and close to a spherical shape. And, indeed, both experiment
and our calculations show a spectra in which the yrast band is more
similar to a linear vibrational one than to a quadratic rotational
one, see Fig.~\ref{fig:Cr-52_spectra}. 
The calculated energies for the yrast band fits well with experiment
up to the expected termination at $I=8\hbar$. In contrast, the model
does not agree with experiment for the yrast states with larger angular
momenta; too high energies are obtained. However, the angular momenta,
which are assigned to those states from experiment, are considered
uncertain. Our calculations support the possibility that these states
have a different angular momenta than the ones they have been attributed.

In experiment, two rotational bands with positive parity and even
spins have been identified. The ground state band up to spin $10\hbar$
and a second band built from the $4_{3}^{+}$-state which goes up
to spin $12\hbar$. And as seen from the transitions in Fig.~\ref{fig:Cr-52_spectra},
our model indeed predicts a rotational band just above the yrast band.
The calculated second band starts at the $0_{2}^{+}$-state and passes
over the $2_{2}^{+}$- and $4_{2}^{+}$-states. For $I\leq6\hbar$
the calculated transitions between the bands are strong. This indicates
a similar internal structure of the two bands for low angular momenta.

\begin{figure}
\includegraphics[clip,width=0.9\columnwidth]{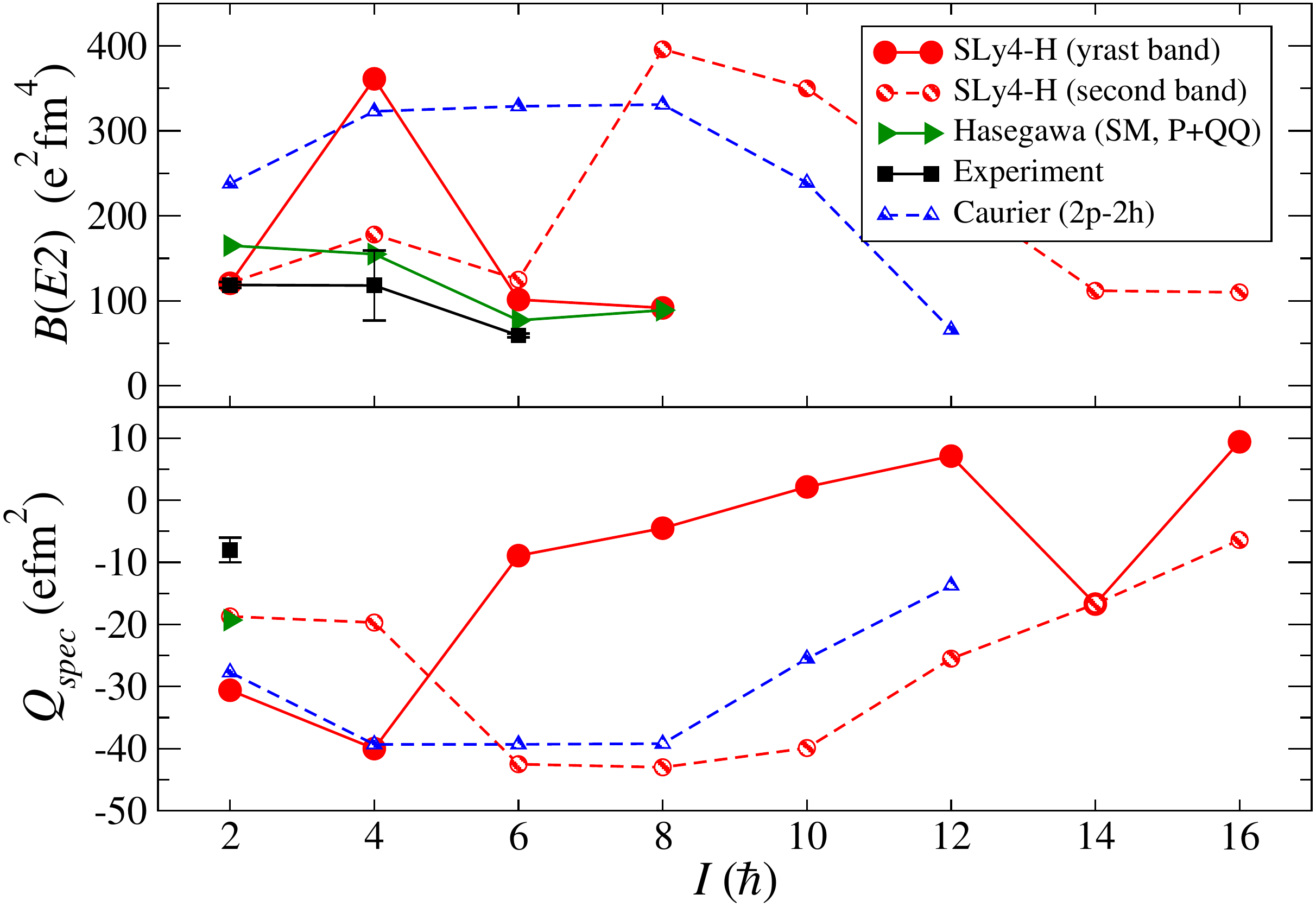}\caption{Transitions and quadrupole moments for $^{52}$Cr. The dashed lines
are for the second band.\label{fig:Cr-52_B_Q}}
\end{figure}

Those two bands, the shell closure yrast band and the excited rotational
band, have been investigated by Caurier et al. \citep{Caurier_2004}.
They found that the yrast band indeed is composed mainly by the closed
shell configuration. In contrast, the excited rotational band is built
upon the $0_{2}^{+}$-state with an internal structure dominated by
two neutrons above the $f_{7/2}$ orbital. In addition, they calculate
$Q_{spec}$ and $B(E2)$ values for the excited band which are compared
in Fig.~\ref{fig:Cr-52_B_Q} with our results. Also, in Fig.~\ref{fig:Cr-52_B_Q},
the results for the yrast band is compared with a SM-calculation \citep{Hasegawa_2002}.
While there is a good overall agreement between theory and experiment
our calculations predict the bands slightly closer in energy and more
mixed than in experiment for $I=0-4$.

\subsection{$^{24}$Mg}

The main reason to test the model on the nucleus $_{12}^{24}$Mg$_{12}$
is to compare the results with the ones presented in references \citep{Bender2008}
and \citep{rodri}. In those references they use a similar procedure
as the one given in this paper. That is, a mean field basis of HFB-states
with different constraints on the $\beta\gamma$-deformations, projections
onto good quantum numbers and mixing using the GCM. 

\begin{figure}
\includegraphics[clip,width=0.9\columnwidth]{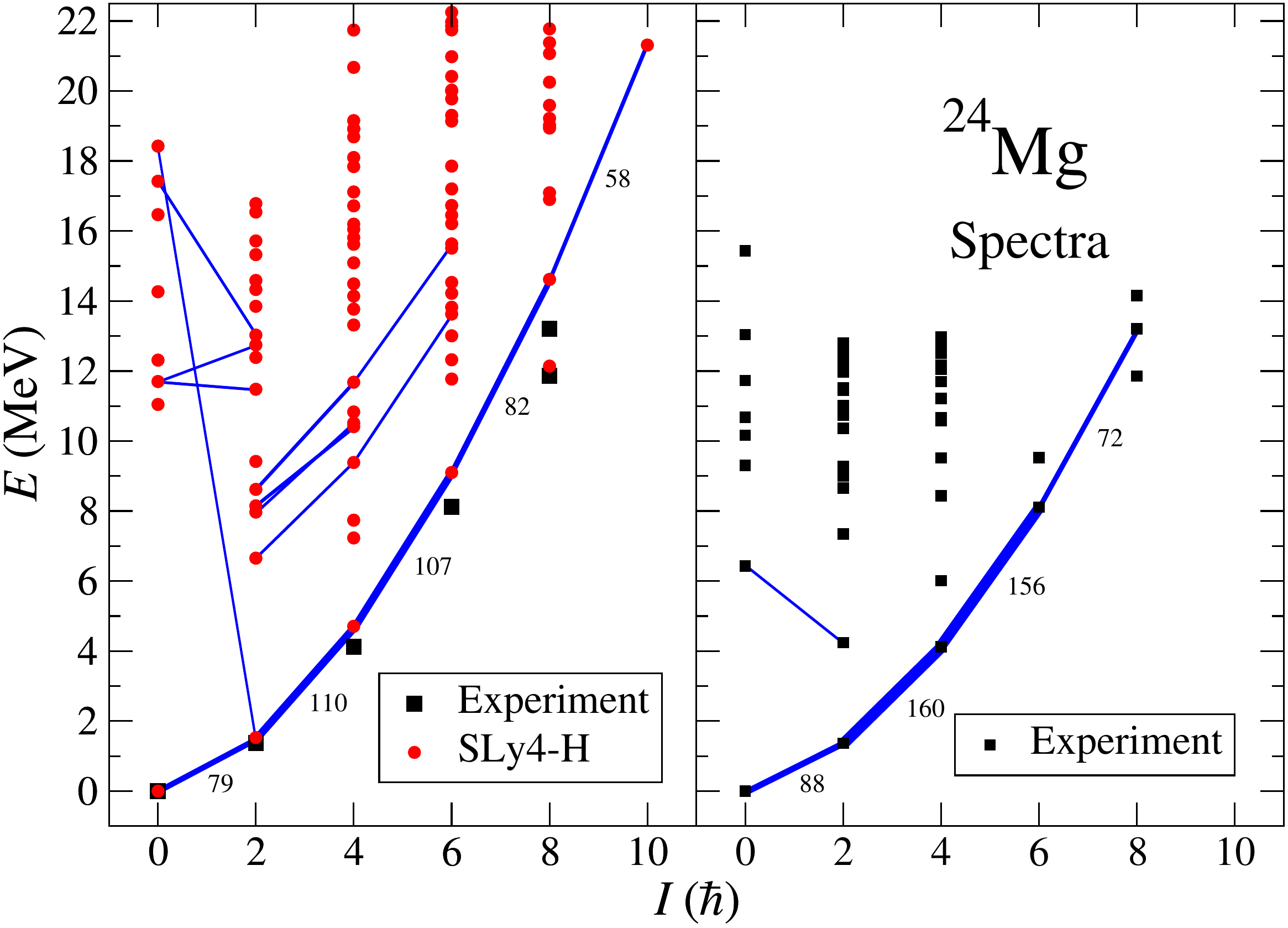}\caption{Same as Fig.~\ref{fig:Cr-48_spectra} for spectra and transitions
in $^{24}$Mg. The experimental values for the $I=8$ states, both
energies and the transitions, are taken from \citep{FIFIELD197877}.\label{fig:Mg_spectra}}
\end{figure}

\begin{figure}
\includegraphics[clip,width=0.9\columnwidth]{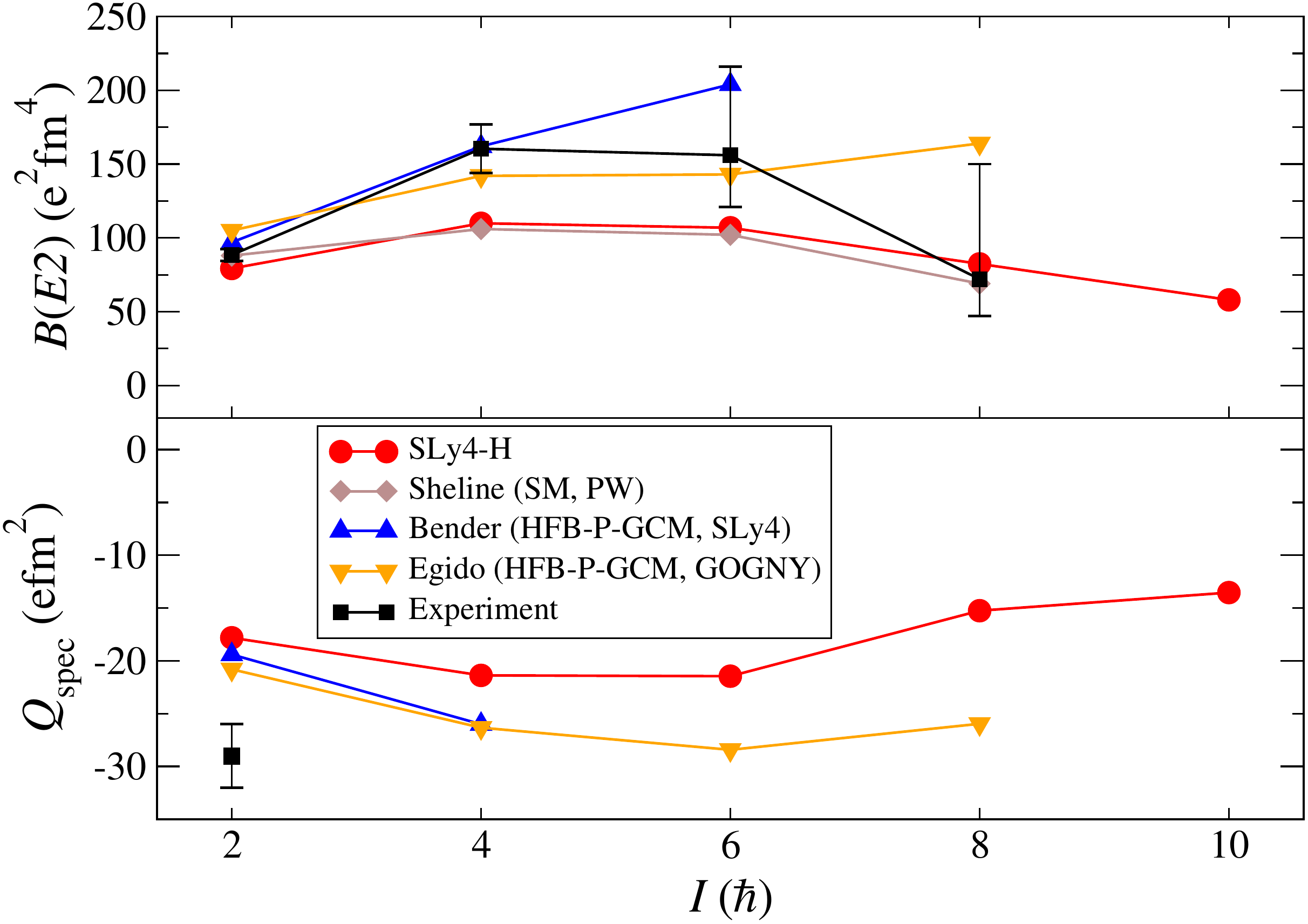}\caption{Transitions and quadrupole moments for $^{24}$Mg.\label{fig:Mg_B_Q}}
\end{figure}

The main difference is that those works use the same interaction throughout
the whole calculation; SLy4 in \citep{Bender2008} and Gogny D1S in
\citep{rodri}. Since those forces are density dependent it is not
well defined how to perform the mixing of states. Therefore, the density
is replaced with the transition density to overcome this problem.
It has been pointed out that this procedure can lead to poles in the
energy for some deformations \citep{poles2009}. This issue is absent
in our model since we postulate a Hamiltonian which can be used in
a straightforward way in the mixing. Our results, together with experiment,
are shown in Fig.~\ref{fig:Mg_spectra} and in Fig. \ref{fig:Mg_B_Q}
$B(E2)$ values and quadrupole moments are compared both with experiments
and the previous calculations. 

The parameters of the calculation for $^{24}$Mg are the same as for
the chromium isotopes except for the following: 10 major shells in
the harmonic oscillator single-particle basis, 300 states to sample
the $\beta\gamma$-plane within $\beta\leq0.8$ and $-30\lyxmathsym{\textdegree}\leq\gamma\leq150^{\lyxmathsym{\textdegree}}$.
Keeping states below 25 MeV in excitation energy resulted in 133 basis
states. The number of points for the projections are 10 for the particle
numbers of both types of nucleons and (6, 12, 24) for the $(\alpha,\beta,\gamma)$
angels of the angular momentum (corresponding to (24, 24, 24) points
in the full space). The cut-off value for the displayed $B(E2)$ transitions
is 7 W.u.

For the spectra, the Gogny force succeeds to reproduce experimental
values in an excellent way, at least up to $I=6$. To the advantage
of our model is that it reproduces the yrast $8_{1}^{+}$ state below
the rotational band; which is seen in experiment. This state is not
reported in \citep{rodri}. 

The $8_{1}^{+}$ state has also been reproduced in SM-calculations
and in the CNS-method; both presented in reference \citep{Sheline_1988}.
In that paper, using the CNS method, it is found that this state is
maximally aligned with its symmetry axis parallel to the axis of rotation.
Hence, it is predicted to be a non-collective state. Indeed, this
is expected to happen at $I=8$ which is the maximal spin that can
be produced for the four valence particles confined to the orbits
of $d_{5/2}$ character. In contrast, the $8_{2}^{+}$ state, which
belongs to the ground state band, is of a collective nature with the
axis of rotation perpendicular to the symmetry axis. This ground state
band continues until it terminates at $I=12$ \citep{Sheline_1988}.
Note the similarity with $^{50}$Cr where the aligned state is yrast
for $I=10$, while the ground band terminates for $I=14$.

Also for the transitions the results from \citep{Bender2008} and
\citep{rodri} fits well with experiment. Again at least up to $I=6$.
But the trend that the $B(E2)$-values increase with higher spin seems
questionable. It is not what one would expect; neither from experiment
nor from experience for rotational bands approaching a terminating
state. Our model agree more with shell model calculations and have
the expected decrease with spin.

\section{Conclusions}

The method introduced works surprisingly well for the description
of both spectra and transitions. Both transitions and delicate structure
information such as backbending are correctly reproduced for all nuclei
considered. Defining a Hamilton operator allows the many-body calculations
to be carried out in a straightforward manner, without any need for
additional assumptions to treat the density-dependent parts of the
functional. In this work, the postulated separable Hamiltonian is
constrained to reproduce the energy surface of a reference EDF. Correctly
describing the detailed landscapes of energy minima and corresponding
shapes is one of the basic components needed in order to reproduce
the experimental spectra.  We have chosen the effective Hamiltonian
as simple as possible, while still capable of producing realistic
results. The simplicity of the interaction, together with the reduction
to the smallest space using the Bloch--Messiah method, allows for
advanced many-body calculations. We have incorporated both collective
and single-particle type excitations using around 200 HFB vacua in
the basis. The number of points needed in the angular momentum projection
increases rapidly when considering higher angular momentum states
\citep{calvin2017} and in this work the calculations have been carried
out to spin 16 while maintaining the refined many-body mixing of the
states. This allowed to cover the spin range up to the terminating
states seen in these nuclei. Thus, to obtain in the complete space
a fully symmetry restored description of the gradual transition from
collective rotation to the non-collective terminating states based
on the GCM.

The present results give encouraging prospects for the future. Any
nuclear interaction can be expressed in terms of sums of separable
terms, through e.g. a singular value decomposition or more refined
physically motivated expansions \citet{ticgai2019,Nesterenko2002}.
Thus, the simple expansion explored here can be fully developed into
a converging expansion. The question in this respect is the applicability
and efficiency of the method. The present study demonstrates the numerical
efficiency of the approach. The effective Hamilton operator employed
here may still be extended with more terms while keeping the calculations
feasible. The first such terms to consider could be an improved pairing
part, hexadecapole terms in the particle-hole part, and a refined
treatment of Coulomb.  

\section*{Acknowledgments}

B.G.C. and J.L. thank the Knut and Alice Wallenberg Foundation (KAW
2015.0021) for financial support. J.R thank the Crafoord foundation
for support. A.I. was supported by Swedish Research Council 2020-03721.
We also acknowledge the Lunarc computing facility. 

\bibliography{biblio}

\appendix

\section{Computation of Matrix elements\label{sec:Appendix:Computation-of-Matrix}}

The Hamiltonian matrix elements in Eq.~(\ref{eq:ham_mat}) can be
expressed \citep{rin80},

\begin{equation}
\left\langle a\left|\hat{H}\right|b\right\rangle =\left\langle a|b\right\rangle \frac{1}{2}\left(\textrm{Tr}\left(\rho\Gamma\right)-\textrm{Tr}\left(\Delta\kappa_{01}^{*}\right)\right).
\end{equation}
The contribution of the two-body interaction in Eq.~(\ref{eq:H_m})
to the particle-hole fields $\Gamma$ becomes

\begin{align}
\Gamma_{ij} & =-\chi\sum_{\mu}\tilde{Q}_{ij}^{2\mu}\textrm{Tr}\left(\rho\left(\tilde{Q}^{2\mu}\right)^{T}\right)\nonumber \\
 & +\chi\sum_{\mu}\left[\tilde{Q}^{2\mu}\rho\left(\tilde{Q}^{2\mu}\right)^{T}\right]_{ij}\nonumber \\
 & +G\left[P\rho^{T}P\right]_{ij}
\end{align}
and the contribution to the particle-particle fields $\Delta$ becomes:

\begin{equation}
\Delta_{ij}=-\frac{1}{2}GP_{ij}\textrm{Tr}\left(\kappa_{10}P^{T}\right)-\chi\left[\tilde{Q}^{2\mu}\kappa_{10}\tilde{Q}^{2\mu}\right]_{ij}.
\end{equation}
In order to speed up the calculations it is important to reduce the
dimensions to the minimal occupied subspace \citep{Yao2009,BONCHE1990}.
We choose a block size which is the maximum value of $n_{a}$ and
$n_{b}$ (see Eq.~(\ref{eq:n-dim})) to denote the size of the upper
left block in the equations below. The BM transformation is used to
transform the $(U,V)$ matrices of the vacua. As an example for the
$\left|b\right\rangle $ vacua we obtain $U_{b}=D_{b}\bar{U}_{b}C_{b}$
and $V_{b}=D_{b}^{*}\bar{V}_{b}C_{b}$ with 
\[
\bar{V}_{b}=\left(\begin{array}{cc}
v_{b} & 0\\
0 & 0
\end{array}\right)\,\,\textrm{and}\,\,\,\bar{U}_{b}=\left(\begin{array}{cc}
u_{b} & 0\\
0 & \mathbb{{1}}
\end{array}\right).
\]

The transitional densities $\rho,\kappa_{10}$ and $\kappa_{01}^{*}$
\citep{rin80} can then be transformed and expressed:

\begin{align}
D_{b}^{\dagger}\rho D_{a}=\bar{\rho} & =\left(\begin{array}{cc}
\bar{\rho}_{11} & 0\\
0 & 0
\end{array}\right)\\
D_{b}\kappa_{10}D_{a}^{T}=\bar{\kappa}_{10} & =\left(\begin{array}{cc}
\left(\bar{\kappa}_{10}\right)_{11} & \left(\bar{\kappa}_{10}\right)_{12}\\
0 & 0
\end{array}\right)\\
D_{b}^{*}\kappa_{01}^{*}D_{a}^{\dagger}=\bar{\kappa}_{01}^{*} & =\left(\begin{array}{cc}
\left(\bar{\kappa}_{01}^{*}\right)_{11} & 0\\
\left(\bar{\kappa}_{01}^{*}\right)_{21} & 0
\end{array}\right)
\end{align}
with

\begin{align}
\bar{\rho}_{11} & =v_{b}^{*}\mathbb{U}^{-1}v_{a}^{T}\\
\left(\bar{\kappa}_{10}\right)_{11} & =v_{b}^{*}\mathbb{U}^{-1}u_{a}^{T}\\
\left(\bar{\kappa}_{10}\right)_{12} & =\left(\kappa_{10}\right)_{11}d_{11}^{-1}d_{12}\\
\left(\bar{\kappa}_{01}^{*}\right)_{11} & =-u_{b}^{*}\mathbb{U}^{-1}v_{a}^{T}\\
\left(\bar{\kappa}_{01}^{*}\right)_{21} & =d_{21}d_{11}^{-1}\left(\bar{\kappa}_{01}^{*}\right)_{11}
\end{align}
where

\begin{equation}
\mathbb{U}^{-1}=\left(v_{a}^{T}d_{11}^{T}v_{b}^{*}+u_{a}^{T}d_{11}^{-1}u_{b}^{*}\right)^{-1}
\end{equation}
and

\begin{equation}
D_{b}^{T}D_{a}^{*}=\left(\begin{array}{cc}
d_{11} & d_{12}\\
d_{21} & d_{22}
\end{array}\right)
\end{equation}
For the blocks of the transformed interaction we introduce the notation:

\begin{align}
\bar{Q}^{2\mu} & =\left(D_{a}^{\dagger}\tilde{Q}^{2\mu}D_{b}\right)_{11}\\
\bar{P}_{1} & =\left(D_{a}^{\dagger}PD_{a}^{*}\right)_{11}\\
\bar{P}_{2} & =\left(D_{b}^{\dagger}PD_{b}^{*}\right)_{11}\\
\bar{P}_{11} & =\left(D_{a}^{\dagger}PD_{b}^{*}\right)_{11}\\
\bar{P}_{12} & =\left(D_{a}^{\dagger}PD_{b}^{*}\right)_{12}\\
\bar{P}_{21} & =\left(D_{a}^{\dagger}PD_{b}^{*}\right)_{21}
\end{align}

If we furthermore decompose the all matrices into proton and neutron
parts and use $q$ to label the proton or neutron blocks we obtain
the full expression for the matrix elements in the optimal space and
with proton and neutron parts explicitly written out as:

\begin{widetext}

\begin{align}
\left\langle a\left|\hat{H}_{Q}+\hat{H}_{P}\right|b\right\rangle  & =\left\langle a|b\right\rangle \frac{1}{2}\left(\sum_{q}\textrm{Tr}\left(\rho^{q}\Gamma^{q}\right)-\sum_{q}\textrm{Tr}\left(\Delta^{q}\kappa_{01}^{*q}\right)\right)\nonumber \\
= & -\left\langle a|b\right\rangle \frac{\chi}{2}\sum_{\mu qq'}\left(-1\right)^{\mu}\textrm{Tr}\left(\bar{\rho}_{11}^{q}\bar{Q}^{2\mu,q}\right)\times\textrm{Tr}\left(\bar{\rho}_{11}^{q'}\bar{Q}^{2\left(-\mu\right),q'}\right)\nonumber \\
 & +\left\langle a|b\right\rangle \frac{\chi}{2}\sum_{q\mu}\left(-1\right)^{\mu}\textrm{Tr}\left(\bar{\rho}_{11}^{q}\bar{Q}^{2\mu,q}\bar{\rho}_{11}^{q}\bar{Q}^{2\left(-\mu\right),q}\right)\nonumber \\
 & +\left\langle a|b\right\rangle \frac{G}{2}\sum_{q}\textrm{Tr}\left(\bar{\rho}_{11}^{q}\bar{P}_{q,1}\left(\bar{\rho}_{11}^{q}\right)^{T}\bar{P}_{q,2}^{*}\right)\nonumber \\
 & -\left\langle a|b\right\rangle \frac{G}{4}\sum_{q}\textrm{Tr}\left(\left[\left(\bar{\kappa}_{10}^{q}\right)_{11},\left(\bar{\kappa}_{10}^{q}\right)_{12}\right]\left[\begin{array}{c}
\bar{P}_{q,11}^{*}\\
\bar{P}_{q,21}^{*}
\end{array}\right]\right)\times\textrm{Tr}\left(\left[\bar{P}_{q,11},\bar{P}_{q,12}\right]\left[\begin{array}{c}
\left(\bar{\kappa}_{01}^{*q}\right)_{11}\\
\left(\bar{\kappa}_{01}^{*q}\right)_{21}
\end{array}\right]\right)\nonumber \\
 & -\left\langle a|b\right\rangle \frac{\chi}{2}\sum_{q}\textrm{Tr}\left(\bar{Q}^{2\mu,q}\left[\left(\bar{\kappa}_{10}^{q}\right)_{11},\left(\bar{\kappa}_{10}^{q}\right)_{12}\right]\left(D_{a}^{\dagger}Q^{2\mu,q}D_{b}\right)^{*}\left[\begin{array}{c}
\left(\bar{\kappa}_{01}^{*q}\right)_{11}\\
\left(\bar{\kappa}_{01}^{*q}\right)_{21}
\end{array}\right]\right)\label{eq:H_full}
\end{align}

\end{widetext}

Where in addition we have used the symmetries of our interaction:

\begin{align*}
Q_{kl}^{2\mu}{}^{*} & =Q_{kl}^{2\mu}\\
Q_{kl}^{2\mu} & =\left(-1\right)^{\mu}Q_{lk}^{2-\mu}\\
P_{kl} & =-P_{lk}\\
P_{kl}^{*} & =P_{kl}
\end{align*}

In the code, this expression is further optimized by moving as many
operations as possible outside the loops of gauge and Euler angles.
The effect of symmetry restoration does not change the canonical occupation
numbers but only leads to a matrix multiplication acting on the $D_{b}$
matrix. 

Expression (\ref{eq:H_full}) becomes the same for HFB states with
odd number parity, the only modification being the application of
the Bloch Messiah decomposition for odd states as described in subsection
\ref{subsec:odd-systems}.
\end{document}